\def\siam{0}  % set to 1 to get the version formatted for siam; delete
\def\arxiv{1} % set to 1 to get the version that arXiv.org is willing to
\ifnum\siam=1
\documentclass[onefignum,onetabnum]{siamltex}
\usepackage{amsmath,epsfig,amsfonts,url,ifthen}

\newcommand{\secref}[1]{\S\ref{sec:#1}}
\newcommand{\appref}[1]{Appendix~\ref{sec:#1}}
\newcommand{\stepref}[1]{Step~\ref{step:#1}}

\newcommand{\thmref}[1]{Theorem~\ref{thm:#1}}
\newcommand{\lemref}[1]{Lemma~\ref{lem:#1}}
\newcommand{\propref}[1]{Proposition~\ref{prop:#1}}
\newcommand{\defref}[1]{Definition~\ref{def:#1}}
\newcommand{\factref}[1]{Fact~\ref{fact:#1}}
\newcommand{\constrref}[1]{Construction~\ref{constr:#1}}

\newtheorem{fact}[theorem]{Fact}
\newtheorem{prop}[theorem]{Proposition}
\newtheorem{defn}{Definition}

\newcounter{theconstr}%[section]
\newenvironment{constr}[1][]
  {\refstepcounter{theconstr}
  {{\sc Construction \thetheconstr}\ifthenelse{\equal{#1}{}}{.}{ (#1).} \ }}
  {}

\else

\documentclass[11pt,letterpaper]{article}
\usepackage{amsmath,amsfonts,amsthm,latexsym,amssymb,color,epsfig}
\usepackage{hyperref,times}
\usepackage{fullpage}
\ifnum\arxiv=0
\usepackage{graphicx}
\fi

\newtheorem{theorem}{Theorem}[section]
\newtheorem{lemma}[theorem]{Lemma}
\newtheorem{prop}[theorem]{Proposition}

\newtheorem{fact}[theorem]{Fact}
\newtheorem{corollary}[theorem]{Corollary}
\theoremstyle{definition}
\newtheorem{definition}{Definition}
\newtheorem{defn}[definition]{Definition}
\newtheorem{constr}{Construction}
\newtheorem{rem}{Remark}
\newtheorem{remark}[rem]{Remark}

\newcommand{\thmref}[1]{Theorem~\ref{thm:#1}}

\newcommand{\lemref}[1]{Lemma~\ref{lem:#1}}
\newcommand{\factref}[1]{Fact~\ref{fact:#1}}
\newcommand{\propref}[1]{Proposition~\ref{prop:#1}}

\newcommand{\defref}[1]{Definition~\ref{def:#1}}

\newcommand{\secref}[1]{Section~\ref{sec:#1}}

\newcommand{\appref}[1]{Appendix~\ref{sec:#1}}
\newcommand{\stepref}[1]{Step~\ref{step:#1}}

\renewcommand{\eqref}[1]{Equation~\ref{eq:#1}}
\newcommand{\constrref}[1]{Construction~\ref{constr:#1}}

\newenvironment{CompactEnumerate}
{
  \vspace{-6pt}
  \begin{list}
      {\arabic{enumi}.}
      {
        \usecounter{enumi}
        \setlength{\leftmargin}{24pt}
        \setlength{\itemsep}{-2pt}
      }
}
{
  \end{list}
  \vspace{-6pt}
}

\newenvironment{CompactItemize}
{
  \vspace{-6pt}
  \begin{list}
     {$\bullet$}
     {
        \usecounter{enumi}
        \setlength{\leftmargin}{24pt}
        \setlength{\itemsep}{-2pt}
     }
}
{
   \end{list}
   \vspace{-6pt}
}

\fi

\newenvironment{myproof}{\begin{proof}%\small
}{\end{proof}}

\newcommand{\tuple}[1]{({#1})}
\newcommand{\set}[1]{\left\{ {#1} \right\}}
\newcommand{\paren}[1]{\left( {#1} \right)}
\newcommand{\floor}[1]{\left\lfloor{#1} \right\rfloor}

\def\bool{\{0,1\}}
\newcommand{\bydef}{\stackrel{\rm def}{=}}
\newcommand{\defeq}{\bydef}

\ifnum\siam=0
\newcommand{\mypar}[1]{\vspace{6pt} \noindent {\sc #1}.\ }
\else
\newcommand{\mypar}[1]{\paragraph{#1}}
\fi

\newcommand{\eps}{\epsilon}

\newcommand{\logeps}{\log\paren{\frac{1}{\eps}}}
\newcommand{\logepstwo}{\log\paren{\frac{1}{\eps_2}}}
\newcommand{\logdel}{\log\paren{\frac{1}{\delta}}}

\newcommand{\bit}[1]{\{0,1\}^{#1}}
\newcommand{\zo}{\{0,1\}}
\newcommand{\from}{\leftarrow}
\renewcommand{\gets}{\leftarrow}
\newcommand{\half}{\frac{1}{2}}

\newcommand{\tm}{{\tilde m}}

\newcommand{\E}{{\mathbb{E}}}
\newcommand{\F}{{\cal F}}
\newcommand{\Hfam}{{\cal H}}

\newcommand{\sd}[1]{\mathbf{SD}\paren{{#1}}}
\newcommand{\SD}[1]{\mathbf{SD}\paren{{#1}}}

\newcommand{\M}{{\cal M}}
\newcommand{\I}{{\cal I}}
\newcommand{\U}{{\cal U}}
\newcommand{\dis}[2]{{\mathsf{dis}(#1,#2)}}
\newcommand{\disfn}{{\mathsf{dis}}}
\newcommand{\syn}{{\mathsf{syn}}}
\ifnum\siam=0 \newcommand{\supp}{{\mathsf{supp}}} \fi
\newcommand{\weight}{{\mathsf{weight}}}

\newcommand{\hinf}{{\mathbf{H}_{\infty}}}
\newcommand{\thinf}{{\tilde{\mathbf{H}}_{\infty}}}
\newcommand{\hepsinf}{{\mathbf{H}_{\infty}^{\eps}}}
\newcommand{\thepsinf}{{\tilde{\mathbf{H}}_{\infty}^{\eps}}}

\ifnum\siam=1

\newcommand{\expe}[2]{\expectation_{{#1}} \left[ {#2} \right]}
\else
\newcommand{\expe}[2]{\E_{{#1}} \left[ {#2} \right]}
\fi

\newcommand{\ff}{\mathsf{SS}}

\newcommand{\loss}{{\lambda}}

\newcommand{\Ext}{\mathsf{Ext}}
\newcommand{\Rec}{\mathsf{Rec}}
\newcommand{\rec}{\mathsf{Rec}}
\newcommand{\Gen}{\mathsf{Gen}}
\newcommand{\gen}{\mathsf{Gen}}
\newcommand{\Rep}{\mathsf{Rep}}
\newcommand{\rep}{\mathsf{Rep}}

\newcommand{\edit}{{\sf Edit}}
\newcommand{\sdif}{{\sf SDif}}

\newcommand{\shin}{{\sf SH}}
\newcommand{\hash}{\mathsf{hash}}

\newcommand{\GF}{\mathit{GF}}

\newcommand{\into}{\ensuremath{\hookrightarrow}}

\DeclareMathOperator{\poly}{poly}

\title{Fuzzy Extractors:
How to Generate Strong Keys from Biometrics and Other Noisy
Data\thanks{A preliminary version of this work appeared in Eurocrypt
2004~\cite{DRS04}.  This version appears in \textit{SIAM Journal on Computing},
38(1):97--139, 2008}}

\author{Yevgeniy Dodis\thanks{{\tt dodis@cs.nyu.edu.}  New York
    University, Department of Computer Science, 251 Mercer St., New
    York, NY 10012 USA. 
\ifnum\siam=1
Phone: (212) 998--3011; 
fax: (212) 995--4124.
\fi
}
\and Rafail Ostrovsky\thanks{{\tt rafail@cs.ucla.edu.} University of
    California, Los Angeles, Department of Computer Science, Box
    951596, 3732D BH, Los Angeles, CA 90095 USA. 
\ifnum\siam=1
Phone: (310) 825--3886;
fax: (310) 825--2273.
\fi
}
\and Leonid Reyzin\thanks{{\tt
    reyzin@cs.bu.edu.}  Boston University, Department of Computer
  Science, 111 Cummington St., Boston MA 02215 USA.
\ifnum\siam=1
Phone: (617) 353--8919;
fax: (617) 353--6457.
\fi
}
\and Adam Smith\thanks{{\tt asmith@cse.psu.edu.} Pennsylvania State
  University, Department of Computer Science and Engineering, 342 IST,
  University Park, PA 16803 USA. 
\ifnum\siam=1
Phone: (814) 865--9505;
fax: (814) 865--3176.
\fi
The research reported here was done
  while the author was a student at the Computer Science and
  Artificial Intelligence Laboratory at MIT and a postdoctoral fellow
  at the Weizmann Institute of Science.}  }

\date{January 20, 2008}

\begin{document}

\ifnum\siam=0
\begin{titlepage}
\def\thepage{}
\fi

\maketitle

\begin{abstract}
\ifnum\siam=0\noindent\fi
We provide formal definitions and efficient secure techniques for
\ifnum\siam=0
\begin{itemize}
\else
\begin{remunerate}
\fi
\ifnum\siam=0\item\else\item[---]\fi
turning noisy information into keys usable for {\em any} cryptographic
application, and, in particular,
\ifnum\siam=0\item\else\item[---]\fi
reliably and securely authenticating biometric data.
\ifnum\siam=0
\end{itemize}
\else
\end{remunerate}
\fi
\ifnum\siam=0\noindent\fi

Our techniques apply not just to biometric information, but to any
keying material that, unlike traditional cryptographic keys, is (1)
not reproducible precisely and (2) not distributed uniformly.  We
propose two primitives: a {\em fuzzy extractor} reliably extracts
nearly uniform randomness $R$ from its input; the
extraction is error-tolerant in the sense that $R$ will be the same
even if the input changes, as long as it remains reasonably close to
the original.  Thus, $R$ can be used as a key in a cryptographic
application.  A {\em secure sketch} produces public information about
its  input $w$ that does not reveal $w$, and yet allows exact
recovery of $w$ given another value that is close to $w$.  Thus, it
can be used to reliably reproduce error-prone biometric inputs without
incurring the security risk inherent in storing them.

We define the primitives to be both formally secure and versatile,
generalizing much prior work.  In addition, we provide nearly optimal
constructions of both primitives for various measures of ``closeness'' of
input data, such as Hamming distance, edit distance, and set difference.

\ifnum\siam=0
\smallskip\noindent{\small \textbf{Key words.}\ \ 
fuzzy extractors, fuzzy fingerprints, randomness extractors,
error-correcting codes, biometric authentication, error-tolerance,
nonuniformity, password-based systems, metric embeddings

}
\fi

\end{abstract}

\ifnum\siam=0
\end{titlepage}
{\small
\tableofcontents
}
\newpage
\fi

\ifnum\siam=1

\pagestyle{myheadings}
\thispagestyle{plain}
\markboth{Y. DODIS, R. OSTROVSKY, L. REYZIN, AND A. SMITH}{FUZZY EXTRACTORS: STRONG
KEYS FROM NOISY DATA}
\fi

\section{Introduction}
\label{sec:intro}

Cryptography traditionally relies on uniformly distributed and precisely
reproducible random strings for its secrets.  Reality, however, makes it
difficult to create, store, and reliably retrieve such strings.  Strings
that are neither uniformly random nor reliably reproducible seem to be more
plentiful.  For example, a random person's fingerprint or iris scan is
clearly not a uniform random string, nor does it get reproduced precisely
each time it is measured.  Similarly, a long pass-phrase (or answers to 15
questions \cite{FJ01} or a list of favorite movies
\cite{JS02}) is not uniformly random and is difficult to remember for
a human user.  This work is about using such nonuniform and
unreliable secrets in cryptographic applications.  Our approach is rigorous
and general, and our results have both theoretical and practical value.

To illustrate the use of random strings
on a simple example, let us consider
the task of password authentication.
A user Alice has a password $w$ and wants to
gain access to her account. A trusted server stores some information
$y=f(w)$ about the password. When Alice enters $w$, the server lets
Alice in only if $f(w)=y$. In this simple application, we assume that
it is safe for Alice to enter the password for the verification.
However, the server's long-term storage is not assumed to be secure (e.g.,
$y$ is stored in a publicly readable  \texttt{/etc/passwd} file in
UNIX~\cite{MT79}).
The goal, then, is to design an efficient $f$ that is hard to invert (i.e.,
given $y$ it is hard to find $w'$ such that $f(w')=y$), so that no one
can figure out Alice's password from $y$. Recall that such
functions $f$ are called {\em one-way functions}.

Unfortunately, the solution above has several problems when used with
passwords $w$ available in real life.  First, the definition of a one-way
function assumes that $w$ is {\em truly uniform} and guarantees nothing if
this is not the case. However, human-generated and biometric passwords are
far from uniform, although they do have some unpredictability in them.
Second, Alice has to reproduce her password
{\em exactly} each time she authenticates herself.  This restriction
severely limits the kinds of passwords that can be used.  Indeed, a human
can precisely memorize and reliably type in only relatively short
passwords, which do not provide an adequate level of security. Greater
levels of security are achieved by longer human-generated and biometric
passwords, such as pass-phrases, answers to questionnaires, handwritten
signatures, fingerprints, retina scans, voice commands, and other values
selected by humans or provided by nature, possibly in combination
(see~\cite{Fryk} for a survey).  These measurements seem to contain much more entropy than human-memorizable passwords.
However, two biometric readings
are rarely identical, even though they are likely to be close;
similarly, humans are unlikely to precisely remember their answers to
multiple questions from time to time, though such answers will likely be
similar.  In other words, the ability to tolerate a (limited) number of
errors in the password while retaining security is crucial if we are to
obtain greater security than provided by typical user-chosen short
passwords.

The password authentication described above is just one example of a
cryptographic application where the issues of nonuniformity and
error-tolerance naturally come up.  Other examples include any
cryptographic application, such as encryption, signatures, or
identification, where the secret key comes in the form of noisy
nonuniform data.

\mypar{Our Definitions} As discussed above, an important general problem is to convert
noisy nonuniform inputs into reliably reproducible,
uniformly random strings.  To this end,
we propose a new primitive, termed \emph{fuzzy
extractor}.  It extracts a uniformly random string $R$ from its input $w$
in a noise-tolerant way.  Noise-tolerance means that if the input changes
to some $w'$ but remains close, the string $R$ can be reproduced exactly.  To
assist in reproducing $R$ from $w'$, the fuzzy extractor outputs a nonsecret
string $P$.  It is important to note that $R$ remains uniformly random even
given $P$.  (Strictly speaking, $R$ will be $\eps$-close to uniform rather
than uniform; $\eps$ can be made exponentially small, which makes $R$
as good as uniform for the usual applications.)

Our approach is general: $R$ extracted from $w$
can be used as a key in a cryptographic application but
unlike traditional keys, need not be stored (because it can be recovered
from any $w'$ that is close to $w$).  We define fuzzy extractors to be {\em
information-theoretically} secure, thus allowing them to be used in
cryptographic systems without introducing additional assumptions
(of course, the cryptographic application itself will typically have
computational, rather than information-theoretic, security).

For a concrete example of how to use fuzzy extractors, in the password
authentication case, the server can store $(P,f(R))$. When the user
inputs $w'$ close to $w$, the server reproduces the actual $R$ using
$P$ and checks if $f(R)$ matches what it stores.  
 The
presence of $P$ will help the adversary invert $f(R)$ only by the additive
amount of $\eps$, because $R$ is $\eps$-close to uniform even given
$P$.\footnote{ To be precise, we should note
that because we do not require $w$, and hence $P$, to be efficiently samplable,
we need $f$ to be a one-way function even in the presence of samples from
$w$; this is implied by security against circuit families.}
Similarly, $R$ can
be used for symmetric encryption, for generating a public-secret key
pair, or for other applications that utilize uniformly random secrets.% 
\footnote{ Naturally, the security of
the resulting system should be properly defined and proven and will
depend on the possible adversarial attacks. In
particular, in this work we do not consider active attacks on $P$ or
scenarios in which the adversary can force multiple invocations of the
extractor with related $w$ and gets to observe the different $P$
values.  See \cite{Boy04,BDKOS05,DKRS06} for follow-up work that
considers attacks on the fuzzy extractor itself.}

\begin{figure}[tbp]
\ifnum\siam=0
\ifnum\arxiv=0
\centerline{\includegraphics{intro-picture}} % usual case -- compiles with either ps or pdf
\else
\centerline{\psfig{file=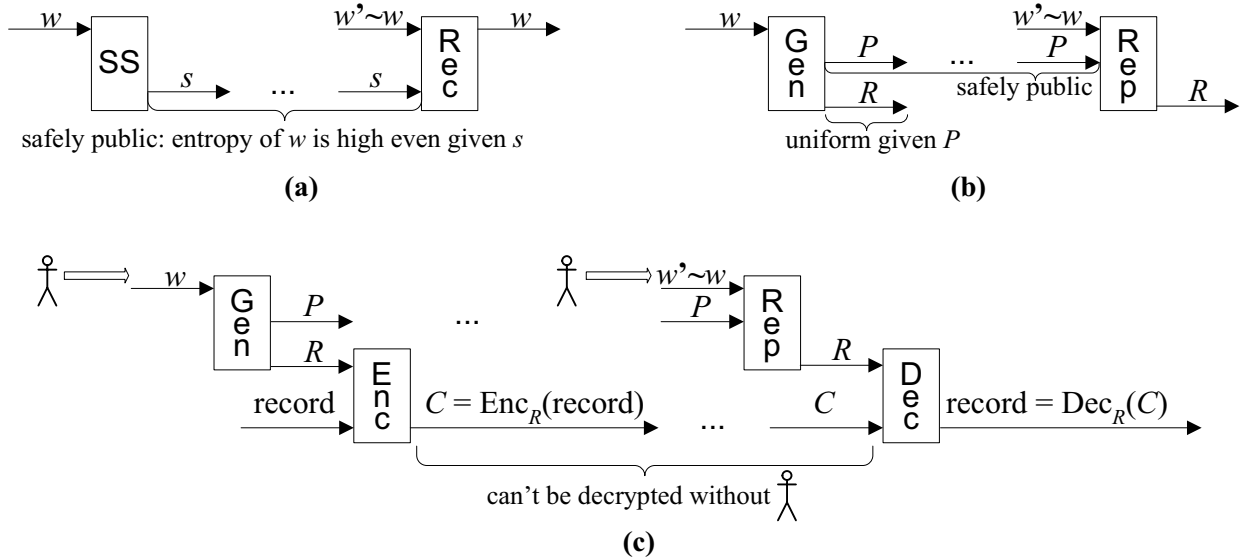}} % arXiv case
\fi
\else
\centerline{\epsfig{file=intro-picture-squished.eps}} % SIAM case
\fi
\caption{\textbf{(a)} secure sketch; \textbf{(b)} fuzzy extractor;
\textbf{(c)} a sample application: user who encrypts a sensitive
record using a cryptographically strong, uniform key $R$ extracted
from biometric $w$ via a fuzzy extractor; both $P$ and the encrypted
record need not be kept secret, because no one can decrypt the record
without a $w'$ that is close.}
\label{figure-fuzzy}
\end{figure}

As a step in constructing fuzzy extractors, and as an interesting
object in its own right, we propose another primitive, termed
\emph{secure sketch}.  It allows precise reconstruction of a noisy
input, as follows: on input $w$, a procedure outputs a sketch $s$.
Then, given $s$ and a value $w'$ close to $w$, it is possible to
recover $w$.  The sketch is secure in the sense that it does not
reveal much about $w$: $w$ retains much of its entropy even if $s$ is
known.  Thus, instead of storing $w$ for fear that later readings will
be noisy, it is possible to store $s$ instead, without compromising
the privacy of $w$.  A secure sketch, unlike a fuzzy extractor, allows
for the precise reproduction of the original input, but does not
address nonuniformity.

Secure sketches, fuzzy extractors and a sample encryption application
are illustrated in Figure~\ref{figure-fuzzy}.

Secure sketches and extractors can be viewed as
providing fuzzy key storage: they allow recovery of the secret key ($w$ or
$R$) from a faulty reading $w'$ of the password $w$ by using some public
information ($s$ or $P$).  In particular, fuzzy extractors can be viewed as
error- and nonuniformity-tolerant secret key {\em key-encapsulation
mechanisms}~\cite{Sho01iso}.

Because different biometric information has different error patterns,
we do not assume any particular notion of closeness between $w'$ and
$w$.  Rather, in defining our primitives, we simply assume that $w$
comes from some metric space, and that $w'$ is no more than a certain
distance from $w$ in that space.  We consider particular metrics
only when building concrete constructions.

\mypar{General Results} Before proceeding to construct our primitives
for concrete metrics, we make some observations about our
definitions. We demonstrate that fuzzy extractors can be built out of
secure sketches by utilizing 
strong {\em randomness
extractors}~\cite{NZ96}, such as, for example, universal
hash functions~\cite{CW79,WC81} (randomness extractors, defined more precisely below,  are families of hash which ``convert'' a high entropy input into a shorter, uniformly distributed output).
We also provide a general technique for constructing secure sketches from
transitive families of isometries,
which is instantiated in concrete constructions later in the paper.
Finally, we
define a notion of a {\em biometric embedding} of one metric space
into another and show that the existence of a fuzzy extractor in the
target space, combined with a biometric embedding of the
source into the target, implies the existence of a fuzzy extractor in the
source space.

These general results help us in building and analyzing our constructions.

\mypar{Our Constructions}
We provide constructions of secure sketches and fuzzy extractors in three
metrics: Hamming distance, set difference, and edit distance. Unless
stated otherwise, all the constructions are new.

Hamming distance (i.e., the number of symbol positions that differ between $w$
and $w'$) is perhaps the most natural metric to consider.  We observe that
the ``fuzzy-commitment'' construction of Juels and Wattenberg~\cite{JW99}
based on error-correcting codes can be viewed as a (nearly optimal)
secure sketch.  We then apply our general result to convert it into a
nearly optimal fuzzy extractor.  While our results on the Hamming distance
essentially use previously known constructions, they serve as an
important stepping stone for the rest of the work.

The set difference metric (i.e., size of the symmetric difference of two input
sets $w$ and $w'$) is appropriate whenever the noisy input is
represented as a subset of features from a universe of
possible features.\footnote{A perhaps unexpected application of the set
difference metric was explored in \cite{JS02}: a user would like to encrypt
a file (e.g., her phone number) using a small subset of values from a large
universe (e.g., her favorite movies) in such a way that those and only
those with a similar subset (e.g., similar taste in movies) can decrypt it.
}
We demonstrate the existence of optimal (with respect to entropy loss)
secure sketches and fuzzy
extractors for this metric.  However, this result is mainly of
theoretical interest, because (1) it relies on optimal constant-weight
codes, which we do not know how to construct, and (2) it produces sketches
of length proportional to the universe size.  We then turn our
attention to more efficient constructions for this metric in order to
handle exponentially large universes.  We provide two such
constructions.

First, we observe that the ``fuzzy vault'' construction of Juels and
Sudan~\cite{JS02} can be viewed as a secure sketch in this metric (and
then converted to a fuzzy extractor using our general result).  We
provide a new, simpler analysis for this construction, which bounds
the entropy lost from $w$ given $s$.  This bound 
is quite high unless one makes the size of the output $s$ very large.  We
then improve the Juels-Sudan construction to reduce the entropy loss and the
length of $s$ to near optimal.  Our improvement in the running time and in
the length of $s$ is exponential for large universe sizes.  However, this
improved Juels-Sudan construction retains a drawback of the original: it
is able to handle only sets of the same fixed size (in particular,
$|w'|$ must equal $|w|$.)

Second, we provide an entirely different construction, called PinSketch, that
maintains the exponential improvements in sketch size and running time and
also handles variable set size.  To obtain it, we note that in the case of
a small universe, a set can be simply encoded as its characteristic vector
(1 if an element is in the set, 0 if it is not), and set difference becomes
Hamming distance.  Even though the length of such a vector becomes
unmanageable as the universe size grows, we demonstrate that this approach
can be made to work quite efficiently even for exponentially large universes (in
particular, because it is not necessary to ever actually write down the
vector).
This involves a result that may be of independent interest: we show
that BCH codes can be decoded in time polynomial in the {\em weight} of the
received corrupted word (i.e., in {\em sublinear} time if the weight is
small).

Finally, edit distance (i.e., the number of insertions and deletions needed
to convert one string into the other) comes up, for example, when the
password is entered as a string, due to typing errors or mistakes made in
handwriting recognition.  We discuss two approaches for secure sketches and
fuzzy extractors for this metric. First, we observe that a recent
low-distortion embedding of Ostrovsky and Rabani~\cite{OR05} immediately
gives a construction for edit distance. The construction performs well when
the number of errors to be corrected is very small (say $n^\alpha$ for
$\alpha<1$) but cannot tolerate a large number of errors. Second, we give a
biometric embedding (which is less demanding than a low-distortion
embedding, but suffices for obtaining fuzzy extractors) from the edit distance
metric into the set difference metric. Composing it with a fuzzy extractor
for set difference gives a different construction for edit
distance, which does better when $t$ is large; it can handle as many as
$O(n/\log^2 n)$ errors with meaningful entropy loss.

Most of the above constructions are quite practical; some implementations
are available~\cite{ijs-bch-impl}.

\mypar{Extending Results for Probabilistic Notions of Correctness}
The definitions and constructions just described use a very strong error
model: we require that secure sketches and fuzzy extractors accept {\em
every} secret $w'$ which is sufficiently close to the original secret $w$,
with probability 1. Such a stringent model is useful, as it makes no
assumptions on the stochastic and computational properties of the error
process.  However, slightly relaxing the error conditions allows
constructions which tolerate a (provably) much larger number of errors, at
the price of restricting the settings in which the constructions can be
applied.  In \secref{improved-param-list}, we extend the definitions and
constructions of earlier sections to several relaxed error models.

It is well-known that in the standard setting of error-correction
for a binary communication channel, one can tolerate many more
errors when the errors are random and independent than when the
errors are determined adversarially.  In contrast, we present fuzzy
extractors that meet Shannon's bounds for correcting random errors
and, moreover, can correct the same number of errors even when
errors are adversarial. In our setting, therefore, under a proper
relaxation of the correctness condition, adversarial errors are no
stronger than random ones. The constructions are quite simple and
draw on existing techniques from the coding
literature~\cite{BBR88,DGL04,Gur03,Lan04,MPSW05}.

\mypar{Relation to Previous Work} Since our work combines elements of
error correction, randomness extraction and password authentication,
there has been a lot of related work.

The need to deal with nonuniform and low-entropy passwords
has long been realized in the security community, and
many approaches have been proposed.
For example, Kelsey et
al.~\cite{KSHW97} suggested using $f(w,r)$ in place of $w$ for the
password authentication scenario, where $r$ is a public random ``salt,''
to make a brute-force attacker's life harder.
While
practically useful, this approach does not add any entropy to the
password and does not formally address the needed properties of $f$.
Another approach, more closely related to ours, is to add biometric
features to the password.
For example, Ellison et al.~\cite{EHMS00} proposed asking the user a
series of $n$ personalized questions and using these answers to encrypt
the ``actual'' truly random secret $R$. A similar approach using the
user's keyboard dynamics (and, subsequently,
voice~\cite{MRLW01a,MRLW01b}) was proposed by Monrose et
al.~\cite{MRW99}. These approaches require the design of
a secure ``fuzzy encryption.''  The above works
proposed  heuristic
designs  (using various forms of
Shamir's secret sharing), but gave no formal analysis.
 Additionally, error tolerance was addressed only by brute force
search.

A formal approach to error tolerance in biometrics was taken by Juels
and Wattenberg~\cite{JW99} (for less formal solutions,
see~\cite{DFMP98,MRW99,EHMS00}), who provided a simple way to tolerate
errors in {\em uniformly distributed} passwords.  Frykholm and
Juels~\cite{FJ01} extended this solution and provided entropy analysis
to which ours is similar.  Similar approaches have been explored
earlier in seemingly unrelated literature on cryptographic information
reconciliation, often in the context of quantum cryptography (where
Alice and Bob wish to derive a secret key from secrets that have small
Hamming distance), particularly~\cite{BBR88,BBCS91}.  Our construction
for the Hamming distance is essentially the same as a component of the
quantum oblivious transfer protocol of~\cite{BBCS91}.

Juels and Sudan~\cite{JS02} provided the first construction for a
metric other than Hamming: they constructed a ``fuzzy vault'' scheme for
the set difference metric.  The main difference is that \cite{JS02}
lacks a cryptographically strong definition of the object constructed.
In particular, their construction leaks a significant amount of
information about their analog of $R$, even though it leaves the
adversary with provably ``many valid choices'' for $R$. In retrospect,
their informal notion is closely related to our secure sketches. Our
constructions in~\secref{setdiff} improve exponentially over the
construction of~\cite{JS02} for storage and computation costs, in the
setting when the set elements come from a large universe.

Linnartz and Tuyls~\cite{LT03} defined and constructed a primitive very
similar to a fuzzy extractor (that line of work was
continued in~\cite{VRDL03}.)  
The definition of \cite{LT03} focuses on the continuous space
$\mathbb{R}^n$ and assumes a particular input distribution (typically
a known, multivariate Gaussian).  Thus, our definition of a fuzzy
extractor can be viewed as a generalization of the notion of a
``shielding function'' from~\cite{LT03}.  However, our constructions
focus on discrete metric spaces.

Other approaches have also been taken for guaranteeing
the privacy of noisy data. Csirmaz and Katona \cite{CK03} considered
quantization for correcting errors in ``physical random functions.''
(This corresponds roughly to secure sketches with no public
storage.)  Barral, Coron and Naccache \cite{BCN04} proposed a system
for offline, private comparison of fingerprints. Although seemingly
similar, the problem they study is complementary to ours, and the two
solutions can be combined to yield systems which enjoy the benefits of
both.

Work on privacy amplification, e.g.,~\cite{BBR88,BBCM95}, as well as work on
derandomization and hardness amplification, e.g.,~\cite{HILL99,NZ96}, also
addressed the need to extract uniform randomness from a random
variable about which some information has been leaked.
A major focus of follow-up
research has been the development of (ordinary, not fuzzy) extractors
with short seeds (see \cite{Sha02} for a survey).
We use extractors in this work (though for our purposes, universal
hashing is sufficient).
Conversely, our work has been applied recently to privacy
amplification: Ding \cite{Din05} used fuzzy extractors for noise
tolerance in Maurer's bounded storage model~\cite{Mau93journal}.

Independently of our work, similar techniques appeared in the
literature on noncryptographic information
reconciliation~\cite{MTZ03,CT04} (where the goal is communication
efficiency rather than secrecy).  The relationship between
secure sketches and efficient information reconciliation is explored
further in~\secref{reconciliation}, which discusses, in particular,
how our secure sketches for set differences provide more efficient
solutions to the set and string reconciliation problems.

\mypar{Follow-up Work} Since the original presentation of this paper
\cite{DRS04}, several follow-up works have appeared (e.g.,
\cite{Boy04,BDKOS05,DS05b,DORS06,Smi07,CL06,LSM06,CFL06}). We refer the
reader to a recent survey about fuzzy extractors \cite{DRS07} for more
information.

\section{Preliminaries}
\label{sec:prelims}

Unless explicitly stated otherwise, all logarithms below are base $2$.
The \emph{Hamming weight} (or just \emph{weight}) of a string is the
number of nonzero characters in it.  We use $U_\ell$ to denote the
uniform distribution on $\ell$-bit binary strings.  If an algorithm
(or a function) $f$ is randomized, we use the semicolon when we wish
to make the randomness explicit: i.e., we denote by $f(x; r)$ the
result of computing $f$ on input $x$ with randomness $r$.  If $X$ is a
probability distribution, then $f(X)$ is the distribution induced on
the image of $f$ by applying the (possibly probabilistic) function
$f$.  If $X$ is a random variable, we will (slightly) abuse notation
and also denote by $X$ the probability distribution on the range of
the variable.

\subsection{Metric Spaces}
A metric space is a set $\M$ with a distance function $\disfn:\M\times \M \to
\mathbb{R}^{+}=[0,\infty)$.  For the purposes of this work,
$\M$ will always be a finite set, and the distance function only take on only
integer values (with $\disfn(x, y) = 0$ if and only if $x=y$) and will
obey symmetry $\disfn(x, y)=\disfn(y, x)$ and the triangle inequality
$\disfn(x, z) \le \disfn(x, y)+\disfn(y,z)$ (we adopt these requirements
for simplicity of exposition, even though the definitions and most of the
results below can be generalized to remove these restrictions).

We will concentrate on the following metrics.
\ifnum\siam=0
\begin{enumerate}
\else
\begin{remunerate}
\fi
\item {\em Hamming metric}. Here $\M = \F^n$ for some alphabet $\F$, and $\dis{w}{w'}$ is the number of positions in
which the strings $w$ and $w'$ differ.
\item {\em Set difference metric}. Here $\M$ consists of all subsets
of a universe $\U$.  For two sets $w, w'$, their symmetric difference
$w\triangle w'\defeq \{x \in w\cup w' \mid x\notin w\cap
w'\}$.  The distance between two sets $w,w'$ is $|w\triangle w'|$.%
~\footnote{In the
preliminary version of this work \cite{DRS04}, we
  worked with this metric scaled by $\half$; that is, the distance was
  $\half | w \triangle w'|$.  Not scaling makes more sense,
  particularly when $w$ and $w'$ are of potentially different sizes
  since
$|w\triangle w'|$ may be odd.  It also agrees with the hamming
  distance of characteristic vectors;
see \secref{setdiff}.}
We will sometimes restrict
$\M$ to contain only $s$-element subsets for some $s$.
\item {\em Edit metric}. Here $\M = \F^*$, and the distance between $w$ and
$w'$ is defined to be the smallest number of character insertions and
deletions needed to transform $w$ into $w'$.%
~\footnote{Again, in  \cite{DRS04}, we
  worked with this metric scaled by $\half$.  Likewise,
  this makes little sense when strings can be of different lengths,
  and we avoid it here.}
(This is different from the
Hamming metric because insertions and deletions shift the characters that
are to the right of the insertion/deletion point.)
\ifnum\siam=0
\end{enumerate}
\else
\end{remunerate}
\fi
\ifnum\siam=1

\fi
As already mentioned, all three metrics seem natural for biometric data.

\subsection{Codes and Syndromes}
Since we want to achieve error tolerance
in various metric spaces, we will use {\em error-correcting codes} for
a particular metric.  A code $C$ is a subset
$\set{w_0,\ldots,w_{K-1}}$ of $K$ elements of $\M$.  The map from $i$
to $w_i$, which we will also sometimes denote by $C$, is called
\emph{encoding}.  The {\em minimum distance} of $C$ is the smallest
$d>0$ such that for all $i\neq j$ we have $\dis{w_i}{w_j}\ge d$. In
our case of integer metrics, this means that one can detect up to
$(d-1)$ ``errors'' in an element of $\M$.  The {\em error-correcting
distance} of $C$ is the largest number $t>0$ such that for every $w\in
\M$ there exists at most one codeword $c$ in the ball of radius $t$
around $w$: $\dis{w}{c} \le t$ for at most one $c\in C$.  This means
that one can correct up to $t$ errors in an element $w$ of $\M$; we
will use the term \emph{decoding} for the map that finds, given $w$,
the $c\in C$ such that $\dis{w}{c}\le t$ (note that for some $w$, such
$c$ may not exist, but if it exists, it will be unique; note also that
decoding is not the inverse of encoding in our terminology).  For
integer metrics by triangle inequality we are guaranteed that $t\ge
\lfloor (d-1)/2\rfloor$.  Since error correction will be more
important than error detection in our applications, we denote the
corresponding codes as $(\M,K,t)$-codes.  For efficiency purposes, we
will often want encoding and decoding to be polynomial-time.

For the Hamming metric over $\F^n$, we will sometimes call $k =
\log_{|\F|} K$ the {\em dimension} of the code and denote the code
itself as an $[n, k, d = 2t+1]_\F$-code, following the standard
notation in the literature. We will denote by $A_{|\F|}(n, d)$ the
maximum $K$ possible in such a code (omitting the subscript when
$|\F|=2$), and by $A(n, d, s)$ the maximum $K$ for such a code over
$\bool^n$ with the additional restriction that all codewords have
exactly $s$ ones.

If the code is linear (i.e., $\F$ is a field,
$\F^n$ is a vector space over $\F$, and $C$ is a linear subspace), then
one can fix a parity-check matrix $H$ as any matrix whose rows
generate the orthogonal space $C^{\perp}$.  Then for any $v\in \F^n$,
the syndrome $\syn(v)\bydef Hv$.  The syndrome of a vector is its
projection onto subspace that is orthogonal to the code and can thus
be intuitively viewed as the vector modulo the code.  Note that
$v\in C \Leftrightarrow\syn(v)=0$.  Note also that $H$ is an $(n-k)\times n$
matrix and that $\syn(v)$ is $n-k$ bits long.

 The syndrome captures all the information necessary for
decoding. That is, suppose a codeword $c$ is sent through a channel
and the word $w=c+e$ is received. First, the syndrome of $w$ is the
syndrome of $e$: $\syn(w)=\syn(c) + \syn(e)=0+ \syn(e) = \syn(e)$.
Moreover, for any value $u$, there is at most one word $e$ of weight
less than $d/2$ such that $\syn(e)=u$ (because the existence of a pair
of distinct words $e_1,e_2$ would mean that $e_1-e_2$ is a codeword of
weight less than $d$, but since $0^n$ is also a codeword and the minimum
distance of the code is $d$, this is impossible). Thus, knowing
syndrome $\syn(w)$ is enough to determine the error pattern $e$ if not
too many errors occurred.

\subsection{Min-Entropy, Statistical Distance, Universal Hashing, and Strong Extractors}\label{sec:worst}
When discussing security, one is often interested in the probability
that the adversary predicts a random value (e.g., guesses a secret
key).  The adversary's best strategy, of course, is to guess the most
likely value.  Thus, \emph{predictability} of a random variable $A$ is
$\max_{a} \Pr[A=a]$, and, correspondingly, {\em min-entropy}
$\hinf(A)$ is $ -\log (\max_{a} \Pr[A=a])$ (min-entropy can thus be
viewed as the ``worst-case'' entropy~\cite{CG88}; see also
\secref{average}).

The min-entropy of a distribution tells us how many nearly uniform random
bits can be extracted from it.  The notion of ``nearly'' is defined as
follows.  The {\em statistical distance between} two probability
distributions $A$ and $B$ is $\sd{A,B} = \frac{1}{2} \sum_v |\Pr(A=v)
-\Pr(B=v)|$.

 Recall the definition of {\em strong randomness
extractors}~\cite{NZ96}.
\begin{defn} \label{def:ext}
Let $\Ext:\bit{n} \to \bit{\ell}$ be a polynomial time probabilistic
function which uses $r$ bits of randomness. We say that
$\Ext$ is an efficient \emph{$(n,m,\ell,\eps)$-strong extractor} if
for all min-entropy $m$ distributions $W$ on $\bit{n}$,
\ifnum\siam=0\( % this mess is needed to deal with an overful hbox in
		% siam style
\else
\[
\fi
\sd{\tuple{\Ext(W;X), X}, \tuple{U_\ell,X}}\le \eps\ifnum\siam=0,\)
\else
\,,\]
\fi where $X$ is
uniform on $\zo^r$.
\end{defn}
\ifnum\siam=1

\fi
Strong extractors can extract at most $\ell = m - 2\logeps
+O(1)$ nearly random bits~\cite{RT00}. Many constructions match
this bound (see Shaltiel's survey~\cite{Sha02} for references).
Extractor constructions are often complex since they seek to minimize
the length of the seed $X$.
For our purposes, the length of $X$ will be less important, so
universal hash functions ~\cite{CW79,WC81} (defined in the lemma below) will
already give us the optimal $\ell = m - 2\logeps + 2$, as given by the
{\em leftover hash lemma} below (see \cite[Lemma 4.8]{HILL99} as well
as references therein for earlier versions):

\begin{lemma}[Universal Hash Functions and the Leftover-Hash /
  Privacy-Ampli\-fication Lemma]
\label{lem:hill}
Assume a family of functions $\{H_x: \{0,1\}^n\to \{0,1\}^\ell\}_{x\in
X}$ is {\em universal}: for all $a\neq b\in\zo^n$, $\Pr_{x\in X}
[H_x(a)=H_x(b)]=2^{-\ell}$. Then, for any random variable
$W$,\footnote{In~\cite{HILL99}, this inequality is formulated in terms
of R\'enyi entropy of order two of $W$; the change to $\hinf(C)$ is
allowed because the latter is no greater than the former.}
\begin{equation}\label{eq:hill}
\SD{\tuple{H_X(W), X}\ ,\ \tuple{U_\ell, X}}\leq \frac{1}{2}\sqrt{2^{-\hinf(W)}
  2^\ell}\,.
\end{equation}
In particular, universal hash functions are
$(n,m,\ell,\eps)$-strong extractors whenever $\ell\leq m-2\logeps+2$.
\end{lemma}

\subsection{Average Min-Entropy}\label{sec:average}

Recall that \emph{predictability} of a random variable $A$ is
$\max_{a} \Pr[A=a]$, and its {\em min-entropy} $\hinf(A)$ is  $ -\log
(\max_{a} \Pr[A=a])$. Consider now a pair of (possibly correlated)
random variables $A,B$.  If the adversary finds out the value $b$ of
$B$, then predictability of $A$ becomes $\max_{a} \Pr[A=a\mid B=b]$.
On average, the adversary's chance of success in predicting $A$ is then
$\expe{b\from{B}} {\max_{a}\Pr[A=a\mid B=b]}$.  Note that we are
taking the \emph{average} over $B$ (which is not under adversarial
control), but the \emph{worst case} over $A$ (because prediction of
$A$ is adversarial once $b$ is known).  Again, it is convenient to
talk about security in log-scale, which is why we define the
\emph{average min-entropy} of $A$ given $B$ as simply the logarithm of
the above:

\[\thinf(A\mid B) \bydef
-\log\paren{\expe{b\from{B}}{\max_{a} \Pr[A=a\mid B=b]}}
= - \log\paren{\expe{b\from{B}}{2^{-\hinf(A\mid
      B=b)}}}\ifnum\siam=0\,.\fi % this period doesn't fit in siam's
				% page width
\]

Because other notions of entropy have been studied in cryptographic
literature, a few words are in order to explain why this definition
is useful.  Note the importance of taking the logarithm
\emph{after} taking the average (in contrast, for instance, to
conditional Shannon entropy). One may think it more natural to
define average min-entropy as $\expe{b\from{B}}{\hinf(A\mid B=b)}$,
thus reversing the order of $\log$ and $\E$.  However, this notion
is unlikely to be useful in a security application.  For a simple
example, consider the case when $A$ and $B$ are 1000-bit strings
distributed as follows: $B = U_{1000}$ and $A$ is equal to the value
$b$ of $B$ if the first bit of $b$ is 0, and $U_{1000}$ (independent
of $B$) otherwise. Then for half of the values of $b$, $\hinf(A\mid
B=b)=0$, while for the other half, $\hinf(A\mid B=b)=1000$, so
$\expe{b\from B}{\hinf(A\mid B=b)} = 500$. However, it would be
obviously incorrect to say that $A$ has 500 bits of security.  In
fact, an adversary who knows the value $b$ of $B$ has a slightly
greater than $50\%$ chance of predicting the value of $A$ by
outputting $b$.  Our definition correctly captures this $50\%$
chance of prediction, because $\thinf(A\mid B)$ is slightly less
than 1. In fact, our definition of average min-entropy is simply the
logarithm of predictability.

The following useful properties of average min-entropy are proven in
\appref{min-prop}. We also refer the reader to \appref{smooth} for a
generalization of average min-entropy and a discussion of the
relationship between this notion and other notions of entropy.

\begin{lemma}\label{lem:hinf}
Let $A,B,C$ be random variables. Then
\ifnum\siam=0
\begin{itemize}
\else
\begin{remunerate}
\fi
\item[(a)] For any $\delta>0$, the
 conditional entropy $\hinf(A|B=b)$ is at least
 $\thinf(A|B)-\log(1/\delta)$ with probability at least $1-\delta$
 over the choice of $b$.
\item[(b)]
If $B$ has at most $2^\loss$ possible values, then
$\thinf(A\mid (B, C)) \geq \thinf((A,B)\mid C) - \loss \geq
\thinf(A\mid C)-\loss$. In particular, $\thinf(A\mid B)
\geq \hinf((A, B))-\loss \geq \hinf(A) - \loss$.
\ifnum\siam=0
\end{itemize}
\else
\end{remunerate}
\fi
\end{lemma}

\subsection{Average-Case Extractors}

Recall from \defref{ext} that a strong extractor allows one to extract
almost all the min-entropy from some nonuniform random variable
$W$. In many situations, $W$ represents the adversary's uncertainty about
some secret $w$ conditioned on some side information $i$. Since this
side information $i$ is often probabilistic, we shall find the
following generalization of a strong extractor useful (see
\lemref{ff}).

\begin{defn} \label{def:ave-ext}
Let $\Ext:\bit{n} \to \bit{\ell}$ be a polynomial time probabilistic
function which uses $r$ bits of randomness. We say that
$\Ext$ is an efficient \emph{average-case} $(n,m,\ell,\eps)$-strong
extractor if for all pairs of random variables $(W,I)$ such that $W$
is an $n$-bit string satisfying $\thinf(W\mid I)\ge m$, we have
$\sd{\tuple{\Ext(W;X), X, I}, \tuple{U_\ell,X,I}}\le \eps$, where $X$
is uniform on $\zo^r$.
\end{defn}

To distinguish the strong extractors
of \defref{ext} from average-case strong extractors,
we will sometimes call the former \emph{worst-case}
strong extractors.  The two notions are closely related,
as can be seen from the following simple application of
\lemref{hinf}(a).

\begin{lemma}\label{lem:lu}
For any $\delta>0$, if $\Ext$ is a (worst-case)
$(n,m-\logdel,\ell,\eps)$-strong extractor, then $\Ext$ is also an
average-case $(n,m,\ell,\eps+\delta)$-strong extractor.
\end{lemma}

\begin{myproof}
Assume $(W,I)$ are such that $\thinf(W\mid I)\ge m$. Let $W_i = (W\mid
I=i)$ and let us call the value $i$ ``bad'' if
$\hinf(W_i)<m-\logdel$. Otherwise, we say that $i$ is ``good''. By
\lemref{hinf}(a), $\Pr(i\mbox{~is~bad})\le \delta$. Also, for any good
$i$, we have that $\Ext$ extracts $\ell$ bits that are $\eps$-close to uniform
from $W_i$. Thus, by conditioning on the ``goodness'' of
$I$, we get
\ifnum\siam=0
\begin{eqnarray*}
\sd{\tuple{\Ext(W;X), X, I}, \tuple{U_\ell,X,I}}&=&
\sum_{i} \Pr(i)\cdot \sd{\tuple{\Ext(W_i;X), X}, \tuple{U_\ell,X}}\\
&\le&
\Pr(i\mbox{~is~bad})\cdot 1 
+ \sum_{\mbox{\tiny{good}~} i} \Pr(i)\cdot
\sd{\tuple{\Ext(W_i;X), X}, \tuple{U_\ell,X}}\\ &\le& \delta+\eps
\end{eqnarray*}
\else
\begin{eqnarray*}
\sd{\tuple{\Ext(W;X), X, I}, \tuple{U_\ell,X,I}}&=&
\sum_{i} \Pr(i)\cdot \sd{\tuple{\Ext(W_i;X), X}, \tuple{U_\ell,X}}\\
&\le&
\Pr(i\mbox{~is~bad})\cdot 1 
\\ & &
+ \sum_{\mbox{\tiny{good}~} i} \Pr(i)\cdot
\sd{\tuple{\Ext(W_i;X), X}, \tuple{U_\ell,X}}\\ &\le& \delta+\eps
\end{eqnarray*}
\qquad
\fi
\end{myproof}

However, for many strong extractors we do not have to suffer this
additional dependence on $\delta$, because the strong extractor
may be already average-case. In particular, this
holds for extractors obtained via universal hashing.

\begin{lemma}[Generalized Leftover Hash Lemma]\label{lem:2wise}
Assume $\{H_x: \{0,1\}^n\to \{0,1\}^\ell\}_{x\in X}$ is a family of
universal hash functions. Then, for any random variables
$W$ and $I$,
\begin{equation}\label{eq:gen-hill}
\SD{\tuple{H_X(W),X,I}\ ,\ \tuple{U_\ell,X,I}}\leq
  \frac{1}{2}\sqrt{2^{-\thinf(W\mid I)} 2^\ell}\,.
\end{equation}
In particular, universal hash functions are {\em average-case}
$(n,m,\ell,\eps)$-strong extractors whenever $\ell\leq m-2\logeps+2$.
\end{lemma}
\begin{myproof}
Let $W_i = (W\mid I=i)$. Then 
\begin{eqnarray*}
\SD{\tuple{H_X(W),X,I}\ ,\ \tuple{U_\ell,X,I}} &=&
\expe{i}{\SD{\tuple{H_X(W_i),X}\ ,\ \tuple{U_\ell,X}}}\\ &\le&
\frac{1}{2}\expe{i}{\sqrt{2^{-\hinf(W_i)} 2^\ell}}\\ &\le&
\frac{1}{2}\sqrt{\expe{i}{2^{-\hinf(W_i)} 2^\ell}}\\
&=& \frac{1}{2}\sqrt{2^{-\thinf(W\mid I)} 2^\ell}\,.
\end{eqnarray*}
In the above derivation, the first inequality follows from the
standard Leftover Hash Lemma (\lemref{hill}), and the second inequality
follows from Jensen's inequality (namely, $\expe{}{\sqrt{Z}} \leq
\sqrt{\expe{}{Z}}$).
\ifnum\siam=1\qquad\fi
\end{myproof}

\section{New Definitions}
\label{sec:defs}

\subsection{Secure Sketches}\label{sec:def-ff}
Let $\M$ be a
metric space with distance function $\disfn$.

\begin{defn} \label{def:ff}
An \emph{$(\M,m,\tm,t)$-secure sketch} is a pair of randomized procedures,
``sketch'' ($\ff$) and ``recover'' ($\Rec$), with the following properties:
\ifnum\siam=0
\begin{enumerate}
\else
\begin{remunerate}
\fi
\item  The sketching procedure $\ff$ on input $w\in \M$
returns a bit string $s\in \bit{*}$.

\item The recovery procedure  $\Rec$ takes an element $w'\in M$ and a bit
string $s\in \bit{*}$.  The \emph{correctness} property of secure sketches
guarantees that if $\dis{w}{w'}\le t$, then $\Rec(w',\ff(w))=w$.  If
$\dis{w}{w'}>t$, then no guarantee is provided about the output of $\Rec$.

\item The \emph{security}
property guarantees that for any distribution $W$ over
$\M$ with min-entropy $m$, the value of $W$ can be recovered by
the adversary who observes $s$ with probability no greater than $2^{-\tm}$.
That is, $\thinf(W\mid \ff(W))\geq \tm$.
\ifnum\siam=0
\end{enumerate}
\else
\end{remunerate}

\fi
A secure sketch is \emph{efficient} if $\ff$ and $\Rec$ run in
expected polynomial time.
\end{defn}

\mypar{Average-Case Secure Sketches} In many situations, it may well
be that the adversary's information $i$ about the password $w$ is
probabilistic, so that sometimes $i$ reveals a lot about $w$, but most
of the time $w$ stays hard to predict even given $i$.  In this case,
the previous definition of secure sketch is hard to apply: it provides
no guarantee if $\hinf(W|i)$ is not fixed to at least $m$ for some bad
(but infrequent) values of $i$.
A more robust definition would provide the same guarantee for all
pairs of variables $(W,I)$ such that predicting the value of $W$ given
the value of $I$ is hard.  We therefore define an \emph{average-case}
secure sketch as follows:

\begin{defn} \label{def:ff-ave}
An {\em average-case $(\M,m,\tm,t)$-secure sketch} is a secure sketch
(as defined in \defref{ff}) whose security property is strengthened as
follows: for any random variables $W$ over $\M$ and $I$ over
$\{0,1\}^*$ such that $\thinf(W\mid I)\ge m$, we have $\thinf(W\mid
(\ff(W), I))\geq \tm$.  Note that an average-case secure sketch is
also a secure sketch (take $I$ to be empty).
\end{defn}

This definition has the advantage that it composes naturally, 
\ifnum\siam=0
as shown in~\lemref{embedff}.  
\else
% to deal with an overfull hbox
as~\lemref{embedff} shows.
\fi
All of our constructions will in fact be
average-case secure sketches.  However, we will often omit the
term ``average-case''  for simplicity of
exposition.

\mypar{Entropy Loss} The quantity $\tm$ is called the {\em residual
  (min-)entropy} of the secure sketch, and the quantity $\loss=m-\tm$
is called the {\em entropy loss} of a secure sketch.  In analyzing the
security of our secure sketch constructions below, we will typically
bound the entropy loss regardless of $m$, thus obtaining families of
secure sketches that work for all $m$ (in  general,
\cite{Rey07} shows that the entropy loss of a secure sketch
is upperbounded by its entropy loss on the uniform distribution of inputs).
Specifically, for a given
construction of $\ff$, $\rec$ and a given value $t$, we will get a
value $\loss$ for the entropy loss, such that, for {\em any $m$},
$(\ff,\rec)$ is an $(\M, m, m-\loss,t)$-secure sketch. In fact, the
most common way to obtain such secure sketches would be to bound the
entropy loss by the length of the secure sketch $\ff(w)$, as given in
the following simple lemma:

\begin{lemma}\label{lem:ent-loss}
Assume some algorithms $\ff$ and $\rec$ satisfy the correctness property of a secure
sketch for some value of $t$, and that the output range of $\ff$
has size at most $2^\loss$ (this holds, in particular, if the length
of the sketch is bounded by $\loss$). Then, for any min-entropy
threshold $m$, $(\ff,\rec)$ form an average-case
$(\M,m,m-\loss,t)$-secure sketch for $\M$. In particular, for any $m$,
the entropy loss of this construction is at most $\loss$.
\end{lemma}
\begin{myproof}
The result follows immediately from \lemref{hinf}(b), since $\ff(W)$ has
at most $2^\loss$ values: for any $(W,I)$, $\thinf(W\mid (\ff(W), I))
\geq \thinf(W\mid I) - \loss$.
\ifnum\siam=1\qquad\fi
\end{myproof}

The above observation formalizes the intuition that a good secure
sketch should be as short as possible. In particular, a short secure
sketch will likely result in a better entropy loss. More discussion
about this relation can be found in \secref{reconciliation}.

\subsection{Fuzzy Extractors}\label{sec:def-fe}

\ifnum\siam=1
\ 

\fi

\begin{defn} \label{def:fe}
An \emph{$(\M,m,\ell,t,\eps)$-fuzzy extractor} is a pair of
randomized procedures, ``generate'' ($\Gen$) and ``reproduce'' ($\Rep$),
with the following properties:

\ifnum\siam=0
\begin{enumerate}
\else
\begin{remunerate}
\fi
\item The generation procedure $\Gen$ on input
$w\in \M$ outputs an extracted string $R\in \zo^\ell$ and a helper
string $P\in \bit{*}$.

\item The reproduction procedure $\Rep$ takes an element $w'\in M$ and
a bit string $P\in \bit{*}$ as inputs.  The \emph{correctness} property of
fuzzy extractors guarantees that if $\dis{w}{w'}\le t$ and $R, P$ were
generated by $(R, P) \from \Gen(w)$, then $\Rep(w',P)=R$.  If
$\dis{w}{w'}>t$, then no guarantee is provided about the output of $\Rep$.

\item
The \emph{security} property guarantees that for any distribution $W$ on
$\M$ of min-entropy $m$, the string $R$ is nearly uniform even
for those who observe $P$: if $\tuple{R,P} \from \Gen(W)$, then
$\sd{\tuple{R,P},\allowbreak \tuple{U_\ell,P} }\allowbreak \le \eps$.

\ifnum\siam=0
\end{enumerate}
\else
\end{remunerate}

\fi
A fuzzy extractor is \emph{efficient} if $\Gen$ and $\Rep$ run
in expected polynomial time.
\end{defn}
\ifnum\siam=1

\fi
In other words, fuzzy extractors allow one to extract some randomness $R$
from $w$ and then successfully reproduce $R$ from any string $w'$ that is
close to $w$.  The reproduction uses the helper string
$P$ produced during the initial extraction; yet $P$ need not remain secret,
because $R$ looks truly random even given $P$. To justify our terminology,
notice that strong extractors (as defined in \secref{prelims}) can indeed
be seen as ``nonfuzzy'' analogs of fuzzy extractors, corresponding to
$t=0$, $P=X$, and $\M = \zo^n$.

We reiterate that the nearly uniform random bits output by a fuzzy
extractor can be used in any cryptographic context that requires
uniform random bits (e.g., for secret keys).  The slight nonuniformity
of the bits may decrease security, but by no more than their distance
$\eps$ from uniform.  By choosing $\eps$ negligibly
small (e.g., $2^{-80}$ should be enough in practice), one can make
the decrease in security irrelevant.

Similarly to secure sketches, the quantity $m-\ell$ is called the
\emph{entropy loss} of a fuzzy extractor.  Also similarly, a more robust
definition is that of an \emph{average-case} fuzzy extractor, which
requires that if $\thinf(W\mid I)\ge m$, then $\sd{\tuple{R,P,I},
\tuple{U_\ell,P,I} }\le \eps$ for any auxiliary random variable $I$.

\section{Metric-Independent Results}

In this section we demonstrate some general results that do not depend
on specific metric spaces.  They will be helpful in obtaining specific
results for particular metric spaces below.  In addition to the
results in this section, 
some generic combinatorial lower bounds on secure sketches and fuzzy
extractors are contained in~\appref{lb}. We will later use these bounds to show the
near-optimality of some of our constructions for the case of uniform
inputs.\footnote{Although we believe our constructions to be near
  optimal for nonuniform inputs as well, and our combinatorial bounds
  in \appref{lb} are also meaningful for such 
inputs, at this time we can use these bounds effectively only for
uniform inputs.} 

\subsection{Construction of Fuzzy Extractors from Secure Sketches}

Not surprisingly, secure sketches are quite useful in constructing
fuzzy extractors. Specifically, we construct fuzzy extractors from
secure sketches and strong extractors as follows: apply $\ff$ to $w$
to obtain $s$, and a strong extractor $\Ext$ with randomness $x$ to
$w$ to obtain $R$.  Store $(s, x)$ as the helper string $P$.  To
reproduce $R$ from $w'$ and $P=(s, x)$, first use $\Rec(w', s)$ to
recover $w$ and then $\Ext(w, x)$ to get $R$.

\medskip
\ifnum\siam=0
\ifnum\arxiv=0
\centerline{\includegraphics{ext-from-sketch}} % usual case -- compiles with either ps or pdf
\else
\centerline{\psfig{file=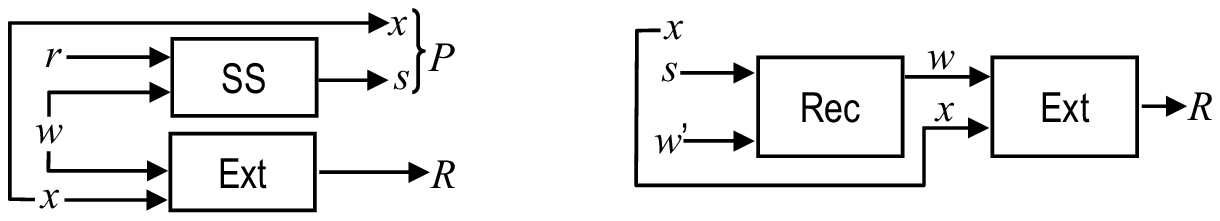}} % arXiv case
\fi
\else
\centerline{\epsfig{file=ext-from-sketch.eps}} % SIAM case
\fi
\smallskip

A few details need to be filled in. First, in order to apply $\Ext$ to
$w$, we will assume that one can represent elements of $\M$ using $n$
bits. Second, since after leaking the secure sketch value $s$, the
password $w$ has only {\em conditional} min-entropy, technically we
need to use the {\em average-case} strong extractor, as defined in
\defref{ave-ext}. The formal statement is given below.

\begin{lemma}[Fuzzy Extractors from Sketches]\label{lem:ff}
  Assume $(\ff, \Rec)$ is an $(\M,m,\allowbreak \tm,\allowbreak t)$-secure sketch, and let
  $\Ext$ be an {\em average-case} $(n,\tm, \ell, \eps)$-strong
  extractor.
Then the following $(\Gen,\Rep)$ is an
  $(\M,m,\ell,t,\eps)$-fuzzy extractor:
\ifnum\siam=0
\begin{itemize}
\else
\begin{remunerate}
\fi
\ifnum\siam=0\item\else\item[---]\fi $\Gen(w;r, x)$: set $P = \tuple{\ff(w;r), x}$, $R =
\Ext(w;x)$, and
 output $\tuple{R,P}$.
\ifnum\siam=0\vspace{0.001ex}\item\else\item[---]\fi $\Rep(w', \tuple{s,x})$: recover $w = \Rec(w',s)$  and output $R
= \Ext(w;x)$.
\ifnum\siam=0
\end{itemize}
\else
\end{remunerate}
\fi
\end{lemma}

\begin{proof}
From the definition of secure sketch (\defref{ff}), we know that
$\thinf(W\mid \ff(W))\ge \tm$. And since $\Ext$ is an average-case
$(n,\tm,\ell,\eps)$-strong extractor,
\ifnum\siam=0
$\SD{(\Ext(W;X),\ff(W),X),(U_\ell,\ff(W),X)} =
\SD{(R,P),(U_\ell,P)}\le \eps$.
\else
\[
\SD{(\Ext(W;X),\ff(W),X),(U_\ell,\ff(W),X)} =
\SD{(R,P),(U_\ell,P)}\le \eps\,.
\]
\qquad
\fi
\end{proof}

On the other hand, if one would like to use a
worst-case strong extractor, we can apply \lemref{lu} to get 

\begin{corollary}
\label{cor:ff-worst-case-ext}
If $(\ff, \Rec)$ is an $(\M,m,\tm,t)$-secure sketch and $\Ext$ is an
$(n,\tm-\logdel, \ell, \eps)$-strong extractor, then the above
construction $(\Gen,\Rep)$ is a $(\M,m,\ell,t,\eps+\delta)$-fuzzy
extractor.
\end{corollary}

Both \lemref{ff} and Corollary~\ref{cor:ff-worst-case-ext} hold
(with the same proofs)
for building \emph{average-case} fuzzy extractors from
\emph{average-case} secure sketches.

While the above statements work for  general extractors, for our purposes we can
simply use universal hashing, since it is an average-case strong
extractor that achieves the optimal
\cite{RT00} entropy loss of $2\logeps$.
In particular, using
\lemref{2wise}, we obtain our main corollary:

\begin{lemma}\label{lem:ff-2wise}
If $(\ff, \Rec)$ is an $(\M,m,\tm,t)$-secure sketch and $\Ext$ is an
$(n,\tm, \ell,\allowbreak \eps)$-strong extractor given by universal hashing (in
particular, any $\ell\le \tm-2\logeps+2$ can be achieved), then the
above construction $(\Gen,\Rep)$ is an $(\M,m,\ell,t,\eps)$-fuzzy
extractor. In particular, one can extract up to $(\tm-2\logeps+2)$
nearly uniform bits from a secure sketch with residual min-entropy
$\tm$.
\end{lemma}

Again, if the above secure sketch is average-case secure, then so is
the resulting fuzzy extractor. In fact, combining the above result
with \lemref{ent-loss}, we get the following general construction of
average-case fuzzy extractors:

\begin{lemma}\label{lem:ext-sketch}
Assume some algorithms $\ff$ and $\rec$ satisfy the correctness property of a secure
sketch for some value of $t$, and that the output range of $\ff$
has size at most $2^\loss$ (this holds, in particular, if the length
of the sketch is bounded by $\loss$). Then, for any min-entropy
threshold $m$, there exists an average-case
$(\M,m,m-\loss-2\logeps+2,t,\eps)$-fuzzy extractor for $\M$. In
particular, for any $m$, the entropy loss of the fuzzy extractor is at
most $\loss+2\logeps-2$.
\end{lemma}

\subsection{Secure Sketches for Transitive Metric Spaces}
\label{sec:trans}
We give a general technique for building secure sketches in {\em
  transitive} metric spaces, which we now define. A permutation $\pi$
on a metric space $\M$ is an \emph{isometry} if it preserves
distances, i.e.,  $\dis{a}{b}=\dis{\pi(a)}{\pi(b)}$.  A family of
permutations $\Pi=\set{\pi_i}_{i\in\I}$ acts {\em transitively} on
$\M$ if for any two elements $a,b\in\M$, there exists $\pi_i\in \Pi$
such that $\pi_i(a)=b$.  Suppose we have a family $\Pi$ of transitive
isometries for $\M$ (we will call such $\M$ {\em transitive}). For
example, in the Hamming space, the set of all shifts $\pi_x(w) = w
\oplus x$ is such a family (see \secref{hamming} for more details on
this example).

\begin{constr}[Secure Sketch For Transitive Metric Spaces]
Let $C$ be an $(\M, K, t)$-code.  Then the general sketching scheme
$\ff$ is the following: given an input $w\in \M$, pick uniformly at
random a codeword $b\in C$, pick uniformly at random a permutation
$\pi\in \Pi$ such that $\pi(w)=b$, and output $\ff(w)=\pi$ (it is
crucial that each $\pi\in \Pi$ should have a canonical description
that is independent of how $\pi$ was chosen and, in particular,
independent of $b$ and $w$; the number of possible outputs of $\ff$
should thus be $|\Pi|$). The recovery procedure $\rec$ to find $w$
given $w'$ and the sketch $\pi$ is as follows: find the closest
codeword $b'$ to $\pi(w')$, and output $\pi^{-1}(b')$.
\end{constr}

Let $\Gamma$
be the number of elements $\pi\in \Pi$ such that $\min_{w,b}
|\{\pi|\pi(w)=b\}| \ge \Gamma$.  I.e., for each $w$ and $b$, there are
at least $\Gamma$ choices for $\pi$.  Then we obtain the following lemma.

\begin{lemma}
\label{lem:transitive}
$(\ff, \rec)$ is an average-case $(\M, m, m-\log|\Pi|+\log \Gamma+\log K,
t)$-secure sketch.  It is efficient if operations on the code, as well
as $\pi$ and $\pi^{-1}$, can be implemented efficiently.
\end{lemma}

\begin{proof}
Correctness is clear: when $\dis{w}{w'}\le t$, then
$\dis{b}{\pi(w')}\le t$, so decoding $\pi(w')$ will result in $b'=b$,
which in turn means that $\pi^{-1}(b')=w$.  The intuitive argument for
security is as follows: we add $\log K+\log \Gamma$ bits of entropy by
choosing $b$ and $\pi$, and subtract $\log|\Pi|$ by publishing $\pi$.
Since given $\pi$, $w$ and $b$ determine each other, the total entropy
loss is $\log|\Pi|-\log K -\log \Gamma$.  More formally, $\thinf(W\mid
\ff(W),I) = \thinf((W, \ff(W))\mid I) - \log|\Pi|$ by \lemref{hinf}(b).  Given a
particular value of $w$, there are $K$ equiprobable choices for $b$
and, further, at least $\Gamma$ equiprobable choices for $\pi$ once $b$ is
picked, and hence any given permutation $\pi$ is chosen with
probability at most $1/(K\Gamma)$ (because different choices for $b$
result in different choices for $\pi$). Therefore, for all  $i$, $w$, and
$\pi$, $\Pr[W=w \wedge \ff(w) = \pi\mid I=i] \le \Pr[W=w\mid I=i]/(K\Gamma)$; hence
$\thinf((W, \ff(W))\mid I) \ge \thinf(W\mid I)+\log K + \log \Gamma$.
\ifnum\siam=1\qquad\fi
\end{proof}

Naturally, security loss will be smaller if the code $C$ is denser.

We will discuss concrete instantiations of this approach
in \secref{hamming} and \secref{permutation-small-set}.

\subsection{Changing Metric Spaces via Biometric Embeddings}
\label{sec:bio-embed}

We now introduce a general technique that allows one to build fuzzy
extractors and secure sketches in some metric space $\M_1$ from fuzzy
extractors and secure sketches in some other metric space $\M_2$. Below, we
let $\dis{\cdot}{\cdot}_i$ denote the distance function in $\M_i$. The
technique is to {\em embed} $\M_1$ into $\M_2$ so as to ``preserve''
relevant parameters for fuzzy extraction.

\begin{defn}
A function $f:\M_1\to \M_2$ is called a $(t_1,t_2,m_1,m_2)$-biometric
embedding if the following two conditions hold:
\ifnum\siam=0
\begin{itemize}
\else
\begin{remunerate}
\fi
\ifnum\siam=0\item\else\item[---]\fi for any $w_1,w_1'\in \M_1$ such that $\dis{w_1}{w_1'}_1\le t_1$,
  we have $\dis{f(w_1)}{f(w_2)}_2\allowbreak \le t_2$.
\ifnum\siam=0\item\else\item[---]\fi for any distribution $W_1$ on $\M_1$ of min-entropy at least
$m_1$, $f(W_1)$ has min-entropy at least $m_2$. 
\ifnum\siam=0
\end{itemize}
\else
\end{remunerate}

\fi
\end{defn}
\ifnum\siam=0\noindent\fi The following lemma is immediate (correctness of the
resulting fuzzy extractor follows from the first condition, and
security follows from the second):
%
%\vspace{-1ex}
\begin{lemma}\label{lem:embedfe}\ifnum\siam=0\sloppypar\fi
If $f$ is a $(t_1,t_2,m_1,m_2)$-biometric embedding of $\M_1$ into
$\M_2$ and $(\gen(\cdot),\rep(\cdot,\cdot))$ is an
$(\M_2,\allowbreak m_2,\allowbreak\ell,\allowbreak t_2,\allowbreak
\eps)$-fuzzy extractor, then $(\gen(f(\cdot)),\allowbreak
\rep(f(\cdot),\cdot))$ is an $(\M_1,\allowbreak m_1,\allowbreak \ell,
\allowbreak t_1, \allowbreak \eps)$-fuzzy extractor.
\end{lemma}
%
%\vspace{-1ex}
\ifnum\siam=0\noindent\fi

It is easy to define \emph{average-case} biometric embeddings (in which
$\thinf(W_1\mid I)\geq m_1 \Rightarrow \thinf(f(W_1)\mid I) \geq m_2$),
which would result in an analogous lemma for average-case fuzzy extractors.

For a similar result to hold for secure sketches, we need biometric
embeddings with an additional property.
\begin{defn}\label{def:embed-ff}
A function $f:\M_1\to \M_2$ is called a $(t_1,t_2,\loss)$-\emph{biometric
embedding with recovery information $g$} if:
\ifnum\siam=0
\begin{itemize}
\else
\begin{remunerate}
\fi
\ifnum\siam=0\item\else\item[---]\fi for any $w_1,w_1'\in \M_1$ such that $\dis{w_1}{w_1'}_1\le t_1$,
  we have $\dis{f(w_1)}{f(w_2)}_2 \allowbreak \le t_2$.
\ifnum\siam=0\item\else\item[---]\fi $g: M_1 \to \zo^*$ is a function with range size at most
$2^\loss$, and $w_1\in M_1$ is uniquely determined by $(f(w_1),
g(w_1))$.
\ifnum\siam=0
\end{itemize}
\else
\end{remunerate}
\fi
\end{defn}

With this definition, we get the following analog of \lemref{embedfe}.
%\vspace{-1ex}
\begin{lemma}\label{lem:embedff}
Let $f$ be a $(t_1,t_2, \loss)$ biometric embedding with recovery
information $g$.  Let $(\ff,\rec)$ be an $(\M_2,\allowbreak m_1-\loss,
\allowbreak\tm_2,\allowbreak t_2)$ average-case secure sketch.  Let
$\ff'(w) = (\ff(f(w)), g(w))$.  Let $\rec'(w', (s, r))$ be the
function obtained by computing $\rec(w', s)$ to get $f(w)$ and then
inverting $(f(w), r)$ to get $w$.  Then $(\ff', \rec')$ is an $(\M_1,
m_1, \tm_2,\allowbreak t_1)$ average-case secure sketch.
\end{lemma}
\begin{myproof}
The correctness of this construction follows immediately from the two
properties given in \defref{embed-ff}. As for security, using
\lemref{hinf}(b) and the fact that the range of $g$ has size at most
$2^\loss$, we get that $\thinf(W\mid g(W)) \ge m_1-\loss$ whenever
$\hinf(W)\ge m_1$. Moreover, since $W$ is uniquely recoverable from
$f(W)$ and $g(W)$, it follows that $\thinf(f(W)\mid g(W)) \ge
m_1-\loss$ as well, whenever $\hinf(W)\ge m_1$. Using the fact that
$(\ff,\rec)$ is an {\em average-case} $(\M_2,\allowbreak m_1-\loss,
\allowbreak\tm_2,\allowbreak t_2)$ secure sketch, we get that
$\thinf(f(W)\mid (\ff(W),g(W))) = \thinf(f(W)\mid \ff'(W))\ge
\tm_2$. Finally, since the application of $f$ can only reduce
min-entropy, $\thinf(W\mid \ff'(W))\ge \tm_2$ whenever $\hinf(W)\ge
m_1$.
\ifnum\siam=1\qquad\fi
\end{myproof}

As we saw, the proof above critically used the notion of average-case
secure sketches. Luckily, all our constructions (for example, those
obtained via \lemref{ent-loss}) are average-case, so this subtlety
will not matter too much.

%
%\vspace{-1ex}
\ifnum\siam=0\noindent\fi

We will see the utility of this novel type of embedding in \secref{edit}.

%\newpage
\section{Constructions for Hamming Distance}
\label{sec:hamming}
In this section we consider constructions for the space $\M=\F^n$
under the Hamming distance metric.  Let $F = |\F|$ and $f=\log_2 F$.

\mypar{Secure Sketches: The Code-Offset Construction}
For the case of $\F=\zo$,
Juels and Wattenberg~\cite{JW99}
considered a notion of ``fuzzy commitment.''
\footnote{In their interpretation, one commits to $x$ by picking
  a random $w$ and publishing $\ff(w;x)$.}  
Given an
$[n,k,2t+1]_2$ error-correcting code $C$ (not necessarily linear),
they fuzzy-commit to $x$ by publishing $w\oplus C(x)$.
Their construction can be rephrased in our language to give a very
simple construction of secure sketches for general $\F$.

We start with an $[n, k, 2t+1]_\F$ error-correcting code $C$ (not
necessarily linear). The idea is to use $C$ to correct errors in $w$
even though $w$ may not be in $C$.  This is accomplished by shifting
the code so that a codeword matches up with $w$, and storing the
shift as the sketch. To do so, we need to view $\F$ as an additive
cyclic group of order $F$ (in the case of most common error-correcting
codes, $\F$ will anyway be a field).

\begin{constr}[Code-Offset Construction]
On input $w$, select a random codeword $c$ (this is
equivalent to choosing a random $x\in \F^k$ and computing $C(x)$), and
set $\ff(w)$ to be the shift needed to get from $c$ to $w$:
$\ff(w)=w-c$.  Then $\rec(w', s)$ is computed by subtracting the shift
$s$ from $w'$ to get $c'=w'-s$; decoding $c'$ to get $c$ (note that
because $\dis{w'}{w}\le t$, so is $\dis{c'}{c}$); and computing $w$ by
shifting back to get $w=c+s$.
\end{constr}

\ifnum\siam=0
\ifnum\arxiv=0
\centerline{\includegraphics{code-offset}} % usual case -- compiles with either ps or pdf
\else
\centerline{\psfig{file=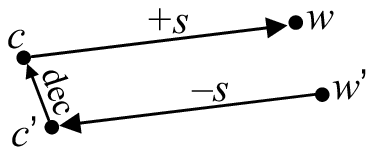}} % arXiv case
\fi
\else
\centerline{\epsfig{file=code-offset.eps}} % SIAM case
\fi
\smallskip

In the case of $\F=\zo$, addition and subtraction are the same, and we
get that computation of the sketch is the same as the Juels-Wattenberg
commitment: $ \ff(w) = w \oplus C(x)$.  In this case, to recover $w$
given $w'$  and $s=\ff(w)$, compute
$c'=w'\oplus s$, decode $c'$ to get $c$, and compute $w=c\oplus s$.

When the code $C$ is linear, this scheme can be simplified as follows.

\begin{constr}[Syndrome Construction]
\label{constr:syndrome}
 Set $\ff(w) = \syn(w)$.  To compute $\rec(w', s)$, find the unique
vector $e\in \F^n$ of Hamming weight $\le t$ such that
$\syn(e)=\syn(w')-s$, and output $w=w'-e$.
\end{constr}

As explained in \secref{prelims}, finding the short error-vector $e$
from its syndrome is the same as
decoding the code.  It is easy to see that
two constructions above are equivalent: given $\syn(w)$ one can
sample from $w-c$ by choosing a random string $v$ with
$\syn(v)=\syn(w)$; conversely, $\syn(w-c)=\syn(w)$.
To show that $\rec$ finds the correct $w$, observe that
$\dis{w'-e}{w'} \le t$ by the constraint on the weight of $e$, and
$\syn(w'-e)=\syn(w')-\syn(e)= \syn(w')-(\syn(w')-s) = s$.  There can
be only one value within distance $t$ of $w'$ whose syndrome is $s$
(else by subtracting two such values we get a codeword that is closer
than $2t+1$ to 0, but 0 is also a codeword), so $w'-e$ must be equal
to $w$.

As mentioned in the introduction, the syndrome construction has
appeared before as a component of some cryptographic protocols over
quantum and other noisy channels~\cite{BBCS91,Cre97}, though it has
not been analyzed the same way.

Both schemes are $(\F^n, m, m-(n-k)f, t)$ secure sketches.  For the
randomized scheme, the intuition for understanding the entropy loss is
as follows: we add $k$ random elements of $\F$ and publish $n$
elements of $\F$.  The formal proof is simply~\lemref{transitive},
because
addition in $\F^n$ is a family of transitive isometries.  For
the syndrome scheme, this follows from~\lemref{ent-loss}, because the
syndrome is $(n-k)$ elements of $\F$.

We thus obtain the following theorem.
\begin{theorem}
\label{thm:hamming-ff}
Given an $[n, k, 2t+1]_\F$ error-correcting code, one can construct
an average-case $(\F^n, m, \allowbreak m-(n-k)f, t)$ secure sketch, which is efficient
if encoding and decoding are efficient.  Furthermore, if the code is
linear, then the sketch is deterministic and its output is $(n-k)$
symbols long.
\end{theorem}

In~\appref{lb} we present some generic lower bounds on secure sketches
and fuzzy extractors.  Recall that $A_F(n,d)$ denotes the maximum
number $K$ of codewords possible in a code of distance $d$ over
$n$-character words from an alphabet of size $F$.  Then by
\lemref{ff-lower}, we obtain that the entropy loss of a secure sketch
for the Hamming metric is at least $nf-\log_2 A_F(n,2t+1)$ when the
input is uniform (that is, when $m=nf$), because $K(\M, t)$ from
\lemref{ff-lower} is in this case equal to $A_F(n, 2t+1)$ (since a
code that corrects $t$ Hamming errors must have minimum distance at
least $2t+1$). This means
that if the underlying code is optimal (i.e., $K=A_F(n,2t+1)$), then the 
code-offset construction above is optimal for the case of uniform
inputs, because its entropy loss is $nf-\log_F K\log_2 F=nf-\log_2 K$.
Of course, we do not know the exact value of $A_F(n,d)$, let alone
efficiently decodable codes which meet the bound, for many settings of
$F$, $n$ and $d$.  Nonetheless, the code-offset scheme gets as close
to optimality as is possible from coding constraints.  If better
efficient codes are invented, then better (i.e., lower loss or higher
error-tolerance) secure sketches will result.

\mypar{Fuzzy Extractors}
As a warm-up, consider the case when $W$ is uniform ($m=n$) and look
at the code-offset sketch construction: $v = w - C(x)$. For $\Gen(w)$,
output $R=x$, $P = v$. For $\rep(w',P)$, decode $w' - P$ to obtain
$C(x)$ and apply $C^{-1}$ to obtain $x$.  The result, quite clearly,
is an $(\F^n,nf,kf,t,0)$-fuzzy extractor, since $v$ is truly random
and independent of $x$ when $w$ is random. In fact, this is exactly the
usage proposed by Juels and Wattenberg~\cite{JW99}, except they viewed
the above fuzzy extractor as a way to use $w$ to ``fuzzy commit'' to
$x$, without revealing information about $x$.

Unfortunately, the above construction setting $R=x$  works only for
uniform $W$, since otherwise $v$ would leak information about
$x$.

In general, we use the construction in \lemref{ff-2wise} combined with
\thmref{hamming-ff} to obtain the following theorem.
%\vspace{-1ex}
\begin{theorem}\label{thm:fe-upper}
  Given any $[n,k,2t+1]_\F$ code $C$ and any $m,\eps$, there exists an
  average-case
  $(\M,m,\ell,\allowbreak t,\eps)$-fuzzy extractor, where $\ell = m +
  kf - nf - 2 \logeps + 2$. The generation $\Gen$ and
  recovery $\Rep$ are efficient if $C$
  has efficient encoding and decoding.
\end{theorem}
%\vspace{-1ex}

\section{Constructions for Set Difference}
\label{sec:setdiff}

We now turn to inputs that are subsets of a universe $\U$;
let $n=|\U|$.
This corresponds to representing an object by a
list of its features. Examples include ``minutiae'' (ridge meetings
and endings) in a fingerprint, short strings which occur in a
long document, or lists of favorite movies.

Recall that the distance between two sets $w,w'$ is the size of their
symmetric difference: $\dis{w}{w'} = |w\triangle w'|$.  We will denote
this metric space by $\sdif(\U)$.  A set $w$ can be viewed as its
\emph{characteristic vector} in $\zo^n$, with $1$ at position $x\in
\U$ if $x\in w$, and $0$ otherwise.  Such representation of sets makes
set difference the same as the Hamming metric. However, we will mostly
focus on settings where $n$ is much larger than the size of $w$,
so that representing a set $w$ by $n$ bits is much less efficient
than, say, writing down a list of elements in  $w$, which requires
only $|w|\log n$ bits.

\mypar{Large Versus Small Universes}
More specifically, we will distinguish two broad categories of
settings.  Let $s$ denote the size of the sets that are given as inputs
to the secure sketch (or fuzzy extractor) algorithms.  Most of this
section studies situations where the universe size $n$ is
superpolynomial in the set size $s$. We call this the ``large
universe'' setting.  In contrast, the ``small universe'' setting
refers to situations in which $n=\mathit{poly}(s)$. We want our
various constructions to run in polynomial time and use polynomial
storage space. In the large universe setting, the $n$-bit string
representation of a set becomes too large to be usable---we will
strive for solutions that are polynomial in $s$ and $\log n$.

In fact, in many applications---for example, when the input is a list of
book titles---it is possible that the actual universe is not only large, but
also difficult to enumerate, making it difficult to even find the position
in the characteristic vector corresponding to $x\in w$.  In that case, it
is natural to enlarge the universe to a well-understood class---for
example, to include all possible strings of a certain length, whether or
not they are actual book titles.  This has the advantage that the position
of $x$ in the characteristic vector is simply $x$ itself; however, because
the universe is now even larger, the dependence of running time on $n$ becomes
even more important.

\mypar{Fixed versus Flexible Set Size}
In some situations, all objects are represented by feature sets of
exactly the same size $s$, while in others the sets may be of
arbitrary size. In particular, the original set $w$ and the corrupted
set $w'$ from which we would like to recover the original need not be
of the same size. We refer to these two settings as {\em fixed} and
{\em flexible} set size, respectively. When the set size is fixed, the
distance $\dis{w}{w'}$ is always even: $\dis{w}{w'}=t$ if and only if
$w$ and $w'$ agree on exactly $s-\frac{t}{2}$ points.  We will denote
the restriction of $\sdif(\U)$ to $s$-element subsets by $\sdif_s(\U)$.

\mypar{Summary}
As a point of reference, we will see below that $\log \binom n s-\log
A(n,2t+1,s)$ is a lower bound on the entropy loss of any secure sketch
for set difference (whether or not the set size is fixed).  Recall that
$A(n,2t+1,s)$ represents the size of the largest code for Hamming
space with minimum distance $2t+1$, in which every word has weight
exactly $s$.  In the large universe setting, where $t \ll n$, the
lower bound is approximately $t\log n$.
The relevant lower bounds are discussed at the end of
Sections~\ref{sec:smallu} and~\ref{sec:ijs}.

In the following sections we will present several schemes which meet this lower bound. The setting of small universes is discussed in \secref{smallu}.
We discuss the code-offset construction (from \secref{hamming}), as well as a permutation-based scheme which is tailored to fixed set size. The latter scheme is optimal for this metric, but impractical.

In the remainder of the section, we discuss schemes for the large universe
setting.  In \secref{ijs} we give an improved version of the
scheme of Juels and Sudan~\cite{JS02}.  Our version achieves optimal
entropy loss and storage $t\log n$ for fixed set size (notice the entropy
loss doesn't depend on the set size $s$, although the running time
does). The new scheme provides an exponential improvement over the original
parameters (which are analyzed in \appref{ojs}). Finally, in
\secref{sublin} we describe how to adapt syndrome decoding algorithms for
BCH codes to our application. The resulting scheme, called PinSketch,
has optimal storage and entropy
loss $t\log (n+1)$, handles flexible set sizes, and is probably the most
practical of the schemes presented here.  Another scheme achieving similar
parameters (but less efficiently) can be adapted from information
reconciliation literature~\cite{MTZ03}; see
\secref{reconciliation} for more details.

We do not discuss fuzzy extractors beyond mentioning here that each
secure sketch presented in this section can be converted to a fuzzy
extractor using \lemref{ff-2wise}.  We have already seen an example of
such conversion in \secref{hamming}.

Table~\ref{table:set-diff} summarizes the constructions discussed in
this section.
\ifnum\siam=0
\begin{table}

{\footnotesize
\begin{tabular}{|c||c|c|c|c||c|}
%  after \\: \hline or \cline{col1-col2} \cline{col3-col4} ...
\hline
                & Entropy Loss     & Storage & Time & Set Size & Notes \\
\hline \hline
  Juels-Sudan   & $t\log n +   \log\left({\binom{n}{r}} / {\binom{n-s}{r-s}}\right) + 2$
                                & $r\log n$ & $poly(r\log(n))$ & Fixed & $r$ is a parameter \\
\cite{JS02} &&&&& $s\leq r \leq n$ \\
\hline
  Generic & $n-\log A(n,2t+1)$ & $n-\log A(n,2t+1)$ & $poly(n)$ & Flexible &  ent. loss $\approx t\log(n)$
  \\
  syndrome && (for linear codes) &&& when $t \ll n$
  \\
\hline
  Permutation-& $\log\binom n s - \log A(n,2t+1,s)$ & $O(n\log n)$ &
  $poly(n)$ & Fixed &ent. loss $\approx t\log n$
\\
based &&&&& when $t \ll n$\\
\hline
  Improved    & $t\log n$     & $t\log n$& $poly(s\log n)$  & Fixed &
  \\
 JS &&&&&
 \\
  \hline
  PinSketch           & $t\log (n+1)$     & $t\log (n+1)$& $poly(s\log n)$  & Flexible  &
  See \secref{sublin}
  \\
  &&&&& for running time\\
\hline
\end{tabular}
}%end \small
\caption{Summary of Secure Sketches for Set Difference.}
\label{table:set-diff}
\end{table}
\else
\begin{table}

{\footnotesize
\begin{tabular}{|c||c|c|c|c|}
%  after \\: \hline or \cline{col1-col2} \cline{col3-col4} ...
\hline
                & Entropy Loss     & Storage & Time & Set Size  \\
\hline \hline
  Juels-Sudan   & $t\log n +   \log\left({\binom{n}{r}} / {\binom{n-s}{r-s}}\right) + 2$
                                & $r\log n$ & $poly(r\log(n))$ & Fixed
                                \\
\cite{JS02}&&&&\\
\hline
  Generic & $n-\log A(n,2t+1)$ & $n-\log A(n,2t+1)$ & $poly(n)$ & Flexible 
  \\
  syndrome && (for linear codes) &&
  \\
\hline
  Permutation-& $\log\binom n s - \log A(n,2t+1,s)$ & $O(n\log n)$ &
  $poly(n)$ & Fixed 
\\
based &&&&\\
\hline
  Improved    & $t\log n$     & $t\log n$& $poly(s\log n)$  & Fixed 
  \\
 JS &&&&
 \\
  \hline
  PinSketch           & $t\log (n+1)$     & $t\log (n+1)$& $poly(s\log n)$  & Flexible  
  \\
  &&&&\\
\hline
\end{tabular}
\\
}%end \small
\caption{Summary of Secure Sketches for Set Difference.  Notes: in
Juels-Sudan scheme, $r$
is a parameter, $s\leq r \leq n$; generic syndrome and
permutation-based schemes achieve entropy loss similar to the improved
Juels-Sudan scheme and PinSketch, $\approx t\log n$, when
$t \ll n$; see \secref{sublin} for running time of PinSketch.}
\label{table:set-diff}
\end{table}
\fi

\subsection{Small Universes}
\label{sec:smallu}

When the universe size is polynomial in $s$, there are a number of natural constructions. The most direct one, given previous work, is the construction of Juels and Sudan \cite{JS02}. Unfortunately, that scheme requires a fixed set size and achieves relatively poor parameters (see \appref{ojs}).

We suggest two possible constructions.  The first involves representing sets as $n$-bit strings and using the constructions of \secref{hamming}.
The second construction, presented below, requires a fixed set size but achieves slightly improved parameters by going through
``constant-weight'' codes.

\mypar{Permutation-based Sketch}
\label{sec:permutation-small-set}
Recall the general construction of \secref{trans} for transitive
metric spaces.  Let $\Pi$ be a set of all permutations on $\U$.  Given
$\pi\in \Pi$, make it a permutation on $\sdif_s(\U)$
naturally: $\pi(w)=\{\pi(x)|x\in w\}$.  This makes $\Pi$ into a family
of transitive isometries on $\sdif_s(\U)$, and thus the
results of \secref{trans} apply.

Let $C\subseteq\bit{n}$ be any $[n,k,2t+1]$ binary code in which all
words have weight exactly $s$. Such codes have been studied
extensively (see, e.g., \cite{AVZ00,BSSS90} for a summary of known
upper and lower bounds). View elements of the code as sets of size
$s$. We obtain the following scheme, which produces a sketch of length
$O(n\log n)$.

\begin{constr}[Permutation-Based Sketch]
On input $w\subseteq\U$ of size $s$,
choose $b\subseteq \U$ at random from the code $C$, and choose
a random permutation $\pi:\U\to\U$ such that $\pi(w)=b$
(that is, choose a random matching between $w$ and $b$ and a
random matching between $\U-w$ and $\U-b$).
Output $\ff(w)=\pi$ (say, by listing $\pi(1),\dots,\pi(n)$).  To recover
$w$ from $w'$ such that $\dis{w}{ w'}\le t$ and $\pi$, compute
$b'=\pi^{-1}(w')$, decode the characteristic vector of $b'$ to obtain
$b$, and output $w=\pi(b)$.
\end{constr}

This construction is efficient as long as decoding is efficient
(everything else takes time $O(n\log n$)).  By \lemref{transitive},
its entropy loss is $\log{\binom{n}{s}}-k$: here $|\Pi|=n!$
and $\Gamma=s!(n-s)!$, so $\log |\Pi| - \log \Gamma =
\log n!/(s!(n-s)!)$.

\mypar{Comparing the Hamming Scheme with the Permutation Scheme}
The code-offset construction was shown to have entropy loss
$n-\log A(n,2t+1)$ if an optimal code is used; the random permutation
scheme has entropy loss $\log{\binom{n}{s}}-\log A(n,2t+1,s)$ for an
optimal code. The Bassalygo-Elias inequality (see \cite{vanLint})
shows that the bound on the random permutation scheme is always at
least as good as the bound on the code offset scheme:
$A(n,d)\cdot 2^{-n} \leq A(n,d,s) \cdot {\binom{n}  s}^{-1}$.
This implies that $n-\log A(n,d) \geq \log{\binom{n}  s}-\log A(n,d,s)$.
Moreover, standard packing arguments give better constructions of
constant-weight codes than they do of ordinary codes.~\footnote{This comes from
  the fact that the intersection of a ball of radius $d$ with the set of all
  words of weight $s$ is much smaller than the ball of radius $d$ itself.}
In fact, the random permutations scheme is optimal for this metric,
just as the code-offset scheme is optimal for the Hamming metric.  

We show this as follows.  Restrict $t$ to be even, because
$\dis{w}{w'}$ is always even if $|w|=|w'|$.  Then the minimum distance
of a code over $\sdif_s(\U)$ that corrects up to $t$ errors must be at
least $2t+1$.%
%\footnote{
Indeed, suppose not.  Then take two codewords, $c_1$ and
$c_2$ such that $\dis{c_1}{c_2} \le 2t$.  There are $k$ elements in
$c_1$ that are not in $c_2$ (call their set $c_1-c_2$) and $k$
elements in $c_2$ that are not in $c_1$ (call their set $c_2-c_1$),
with $k\le t$.  Starting with $c_1$, remove $t/2$ elements of
$c_1-c_2$ and add $t/2$ elements of $c_2-c_1$ to obtain a set $w$
(note that here we are using that $t$ is even; if $k<t/2$, then use
$k$ elements).  Then $\dis{c_1}{w} \le t$ and $\dis{c_2}{w} \le t$,
and so if the received word is $w$, the receiver cannot be certain whether
the sent word was $c_1$ or $c_2$ and hence cannot correct $t$ errors.
%}

Therefore by
\lemref{ff-lower},
we get that the entropy loss of a secure sketch must be at least
$\log{\binom{n}  s} - \log A(n,2t+1,s)$ in the case of a uniform
input $w$.
Thus in principle, it is better to use the random permutation scheme.
Nonetheless, there are caveats.  First, we do not know of
\emph{explicitly} constructed constant-weight codes that beat the
Elias-Bassalygo inequality and would thus lead to better entropy loss
for the random permutation scheme than for the Hamming scheme (see
\cite{BSSS90} for more on constructions of constant-weight codes and
\cite{AVZ00} for upper bounds).  Second, much more is known about
efficient implementation of decoding for ordinary codes than for
constant-weight codes; for example, one can find off-the-shelf
hardware and software for decoding many binary codes. In practice, the
Hamming-based scheme is likely to be more useful.

\subsection{Improving the Construction of Juels and Sudan}
\label{sec:ijs}

We now turn to the large universe setting, where $n$ is
superpolynomial in the set size $s$, and we would like operations
to be polynomial in $s$ and $\log n$.

Juels and Sudan \cite{JS02} proposed a secure sketch for the set
difference metric with fixed set size (called a ``fuzzy vault'' in
that paper). We present their original scheme here with an
analysis of the entropy loss in \appref{ojs}. In particular,
our analysis shows that the original scheme
has good entropy loss only when the storage space is very large.

We suggest an improved version of the Juels-Sudan scheme which is simpler and
achieves much better parameters. The entropy loss and storage space
of the new scheme are both $t\log n$, which is optimal. (The same
parameters are also achieved by the BCH-based construction PinSketch in
\secref{sublin}.)
Our scheme has the advantage of being even simpler to analyze, and the
computations are simpler. As with the original Juels-Sudan scheme, we
assume $n=|\U|$ is a prime power and work over $\F=\GF(n)$.

An intuition for the scheme is that the numbers $y_{s+1},\dots,y_r$
from the JS scheme need not be chosen at random. One can instead
evaluate them as $y_i=p'(x_i)$ for some polynomial $p'$. One can
then represent the entire list of pairs $(x_i,y_i)$ implicitly,
using only a few of the coefficients of $p'$. The new sketch is
deterministic (this was not the case for our  preliminary version
in \cite{DRS04}).  Its implementation is available~\cite{ijs-bch-impl}.

\newcommand{\phigh}{p_{\mathrm{high}}}
\newcommand{\plow}{p_{\mathrm{low}}}

\begin{constr}[Improved JS Secure Sketch for Sets of Size $s$]
\label{constr:ijs}\ifnum\siam=0 \ \\\fi
To compute $\ff(w)$:
\ifnum\siam=0
\begin{CompactEnumerate}
\else
\begin{remunerate}
\fi
\item Let $p'()$ be the unique monic polynomial of degree exactly $s$ such that
  $p'(x) = 0$ for all $x\in w$.  \ifnum\siam=0\\\fi
  (That is, let $p'(z) \defeq \prod_{x \in w} (z-x)$.)
\item Output the coefficients of $p'()$ of degree $s-1$ down to $s-t$. \\
This is equivalent to computing and outputting the first $t$
symmetric polynomials of the values in $A$; i.e., if $w =
\set{x_1,\dots,x_s}$, then output
$$\sum_{i} x_i, \ \sum_{i\neq j} x_i x_j,\ \ldots,\ \sum_{S\subseteq [s],
|S|=t}\paren{\prod_{i \in S} x_i}.$$
\ifnum\siam=0
\end{CompactEnumerate}
\else
\end{remunerate}

\fi
To compute $\rec(w', p')$, where $w'=\{a_1, a_2, \dots, a_s\}$,
\ifnum\siam=0
\begin{CompactEnumerate}
\else
\begin{remunerate}
\fi
\item Create a new polynomial $\phigh$, of degree $s$ which shares
the top $t+1$ coefficients of $p'$; that is, let $\phigh(z) \defeq
z^s + \sum_{i=s-t}^{s-1} a_iz^i$.
\item Evaluate $\phigh$ on all points in $w'$ to obtain $s$ pairs
  $(a_i, b_i)$.
\item\label{step:rsd} Use $[s,s-t,t+1]_\U$ Reed-Solomon decoding
(see, e.g.,~\cite{Blahut83,vanLint})
 to search
for a polynomial $\plow$ of
degree $s-t-1$ such that $\plow(a_i)=b_i$ for at least $s-t/2$ of the
$a_i$ values.
If no such polynomial exists, then stop and output ``fail.''
\item Output the list of zeroes (roots) of the polynomial $\phigh -
  \plow$ (see, e.g.,~\cite{Sho05} for root-finding algorithms; they can be
sped up by first factoring out the known roots---namely, $(z-a_i)$ for
the $s-t/2$ values of $a_i$ that were  not deemed erroneous in the previous
step). \\
\ifnum\siam=0
\end{CompactEnumerate}
\else
\end{remunerate}
\fi
\end{constr}

To see that this secure sketch can tolerate $t$ set difference
errors, suppose  \ifnum\siam=1 that \fi $\dis{w}{w'}\leq t$. Let
$p'$ be as in the sketch algorithm; that is, $p'(z) = \prod_{x \in w}
(z-x)$. The polynomial $p'$ is monic; that is, its leading term is
$z^s$. We can divide the remaining coefficients into two groups: the
high coefficients, denoted $a_{s-t},\dots,a_{s-1}$, and the low
coefficients, denoted
$b_1,\dots,b_{s-t-1}$:
\[
p'(z) = \qquad \underbrace{z^s + \sum_{i=s-t}^{s-1} a_i
  z^i}_{\phigh(z)}\qquad + \qquad \underbrace{\sum_{i=0}^{s-t-1} b_i
  z^i}_{q(z)} \,.
\]
We can write $p'$ as $\phigh + q$, where $q$ has degree $s-t-1$. The
recovery algorithm gets the coefficients of $\phigh$ as input. For any
point $x$ in $w$, we have $0=p'(x)=\phigh(x) + q(x)$. Thus,
$\phigh$ and $-q$ agree at all points in $w$. Since the set $w$
intersects $w'$ in at least $s-t/2$ points, the polynomial $-q$
satisfies the conditions of \stepref{rsd} in $\rec$. That
polynomial is unique, since no two distinct polynomials of degree
$s-t-1$ can get the correct $b_i$ on more than $s-t/2$ $a_i$s (else,
they agree on at least $s-t$ points, which is impossible).
Therefore, the recovered polynomial $\plow$ must be $-q$; hence
$\phigh(x)-\plow(x)=p'(x)$.  Thus, $\rec$ computes the correct $p'$
and therefore correctly finds the set $w$, which consists of the roots
of $p'$.

Since the output of $\ff$ is $t$ field elements, the entropy loss of
the scheme is at most $t\log n$ by \lemref{ent-loss}.  (We will see below that this bound is tight, since any sketch must lose at least $t\log n$ in some situations.) We have proved:

\begin{theorem}[Analysis of Improved JS]\label{lem:js-nice}
    \constrref{ijs} is an average-case $(\sdif_s(\U), m,m-t\log n,t)$ secure
    sketch. The entropy loss and storage of the scheme are at most $t\log n$, and both the sketch generation $\ff()$ and the recovery procedure $\rec()$ run in time polynomial in $s$, $t$ and $\log n$.
\end{theorem}

\mypar{Lower Bounds for Fixed Set Size in a Large Universe}
The short length of the sketch makes this scheme feasible for
essentially any ratio of set size to universe size (we only need
$\log n$ to be polynomial in $s$). Moreover, for large universes the
entropy loss $t\log n$ is essentially optimal for uniform inputs
(i.e., when
$m=\log{\binom{n}  s}$).  We show this as follows.
As already mentioned in the \secref{permutation-small-set},
\lemref{ff-lower} shows that for a uniformly distributed input, the best
possible entropy loss is $m-m'\geq \log {\binom{n} s}-\log
A(n,2t+1,s)$.  

By Theorem 12 of Agrell  {\sl et al.} \cite{AVZ00}, 
\ifnum\siam=0
$A(n,2t+2,s) \leq \frac{\binom n {s-t}}{\binom s {{s-t}}}$.
\else
\[A(n,2t+2,s) \leq \frac{\binom n {s-t}}{\binom s {{s-t}}}\,.\]
\fi
Noting that $A(n, 2t+1, s)=A(n, 2t+2, s)$ because
distances in $\sdif_s(\U)$ are even, the entropy loss is at least
%\vspace{-1ex}
\ifnum\siam=0
$$
m-m' \geq \log {\binom{n}  s} - \log A(n,2t+1,s) \geq
\log {\binom{n}  s} -\log\paren{ {\binom{n}  {s-t}}
\Big / {\binom s {s-t}} } = 
\log {\binom{n-s+t}{t}}\,.$$ 
\else
\begin{eqnarray*}
m-m' & \geq &  \log {\binom{n}  s} - \log A(n,2t+1,s)\\
& \geq &
\log {\binom{n}  s} -\log\paren{ {\binom{n}  {s-t}}
\Big / {\binom s {s-t}} }\\
&  = & 
\log {\binom{n-s+t}{t}}\,.
\end{eqnarray*}
\fi

When $n\gg s$, this last quantity is
roughly $t\log n$, as desired.

\newcommand{\erc}{t}
\newcommand{\te}{m}

\subsection{Large Universes via the Hamming Metric: Sublinear-Time Decoding}
\label{sec:sublin}

In this section, we show that the syndrome construction of
\secref{hamming} can in fact be adapted for small sets in a large
universe, using specific properties of algebraic codes. We will show
that BCH codes, which contain Hamming and Reed-Solomon codes as
special cases, have these properties. As opposed to the constructions
of the previous section, the construction of this section is flexible
and can accept input sets of any size.

Thus we obtain a sketch for sets of flexible size, with entropy loss
and storage $t\log (n+1)$.  We will assume that $n$ is one less than a
power of 2: $n=2^m-1$ for some integer $m$, and will identify $\U$
with the nonzero elements of the binary finite field of degree $m$:
$\U=\GF(2^m)^*$.

\mypar{Syndrome Manipulation for Small-Weight Words}
Suppose now that we have a small set $w\subseteq \U$ of size $s$,
where $n \gg s$.  Let $x_w$ denote the characteristic vector of $w$
(see the beginning of \secref{setdiff}).  Then the syndrome
construction says that $\ff(w) = \syn(x_w)$.  This is an $(n-k)$-bit
quantity.  Note that the syndrome construction gives us no special
advantage over the code-offset construction when the universe is
small: storing the $n$-bit $x_w+C(r)$ for a random $k$-bit $r$ is not
a problem. However, it's a substantial improvement when $n \gg n-k$.

If we want to use $\syn(x_w)$ as the sketch of $w$, then we must
choose a code with $n-k$ very small. In particular, the entropy of $w$
is at most $\log{\binom{n} s}\approx s \log n$, and so the entropy
loss $n-k$ had better be at most $s\log n$.   Binary BCH codes are
suitable for our purposes: they are
 a family of $[n,k,\delta]_2$ linear codes with $\delta=2\erc+1$
and $k = n-\erc m$ (assuming $n=2^m-1$) (see, e.g.
\cite{vanLint}). These codes are optimal for $\erc\ll n$ by the Hamming
bound, which implies that $k\leq n - \log{\binom{n}  \erc}$
\cite{vanLint}.%
\footnote{The Hamming bound is based on the observation that for any code of distance $\delta$, the balls of radius $\floor{(\delta-1)/2}$ centered at various codewords must be disjoint. Each such ball contains ${\binom{n}{\floor{(\delta-1)/2}}}$ points, and so $2^k{\binom{n}{\floor{(\delta-1)/2}}} \leq 2^n$. In our case $\delta=2\erc+1$, and so the bound yields $k\leq n - \log{\binom{n}{\erc}}$.} %
Using the syndrome sketch with a BCH code
$C$, we get entropy loss $n-k=\erc\log (n+1)$, essentially the same as
the $t\log n$ of the
improved Juels-Sudan scheme (recall that $\delta\geq 2\erc+1$ allows us to
correct $\erc$ set difference errors).

The only problem is that the scheme appears to require computation
time $\Omega(n)$, since we must compute $\syn(x_w)=Hx_w$ and, later,
run a decoding algorithm to recover $x_w$. For BCH codes, this
difficulty can be overcome. A word of small weight $w$ can be
described by listing the positions on which it is nonzero.  We call
this description the {\em support} of $x_w$ and write $\supp(x_w)$
(note that  $\supp(x_w)=w$; see the discussion of enlarging the universe
appropriately at the beginning of \secref{setdiff}).

The following lemma holds for general BCH codes (which include binary BCH codes and Reed-Solomon codes as special cases). We state it for binary codes since that is most relevant to the application:

\begin{lemma}\label{lem:bch}
  For a $[n,k,\delta]$ binary BCH code $C$ one can compute:
  \ifnum\siam=0
  \begin{CompactItemize}
  \else
  \begin{remunerate}
  \fi
    \ifnum\siam=0\item\else\item[---]\fi $\syn(x)$, given $\supp(x)$, in time polynomial in $\delta$,
    $\log n$, and $|\supp(x)|$
    \ifnum\siam=0\item\else\item[---]\fi $\supp(x)$, given $\syn(x)$ (when $x$ has weight at most
      $(\delta-1)/2$), in time polynomial in $\delta$ and $\log n$.
  \ifnum\siam=0
  \end{CompactItemize}
  \else
  \end{remunerate}
  \fi
\end{lemma}

The proof of \lemref{bch}  requires a careful reworking of the
standard BCH decoding algorithm. The details are presented in
 \appref{bch}.
For now, we present the
resulting secure sketch for set difference.

\begin{constr}[PinSketch]
\label{constr:bch}\ifnum\siam=0 \ \\ \fi
To compute $\ff(w) = \syn(x_w)$:
\ifnum\siam=0
\begin{CompactEnumerate}
\else
\begin{remunerate}
\fi
    \item Let $s_i=\sum_{x\in w} x^i$   (computations
      in $\GF(2^m)$).
    \item Output $\ff(w) = (s_1,
      s_3,s_5,\dots,s_{2\erc-1})$.
\ifnum\siam=0
\end{CompactEnumerate}
\else
\end{remunerate}
\fi
To recover $\rec(w', (s_1, s_3, \dots, s_{2\erc-1}))$:
\ifnum\siam=0
\begin{CompactEnumerate}
\else
\begin{remunerate}
\fi
  \item Compute $(s'_1, s'_3, \dots, s'_{2\erc-1}) =\ff(w')=\syn(x_{w'})$.
  \item Let $\sigma_i=s'_i-s_i$ (in $\GF(2^m)$, so ``$-$'' is the same as ``$+$'').
  \item Compute $\supp(v)$ such that $\syn(v)=(\sigma_1, \sigma_3, \dots,
  \sigma_{2\erc-1})$ and $|\supp(v)| \le t$ by \lemref{bch}.
  \item If $\dis{w}{w'}\le t$, then $\supp(v) = w\triangle w'$.
  Thus, output $w=w'\triangle \supp(v)$.
\ifnum\siam=0
\end{CompactEnumerate}
\else
\end{remunerate}
\fi
\end{constr}

\ifnum\siam=0\noindent\fi
An implementation of this construction, including the
reworked BCH decoding algorithm, is available~\cite{ijs-bch-impl}.

The bound on entropy loss is easy to see: the output is
$\erc\log (n+1)$ bits long, and hence the entropy loss is at most $\erc\log
(n+1)$ by \lemref{ent-loss}. We obtain:

\begin{theorem}\label{thm:bch}
  PinSketch  is an average-case $(\sdif(\U), \te,\te-\erc\log (n+1),t)$
  secure sketch for set difference with storage $\erc\log (n+1)$. The
  algorithms $\ff$ and $\Rec$ both run in time polynomial in $t$ and
  $\log n$.
\end{theorem}

\section{Constructions for Edit Distance}\label{sec:edit}

The space of interest in this section is the space $\F^*$ for some alphabet
$\F$, with distance between two strings defined as the number of character
insertions and deletions needed to get from one string to the
other.  Denote this space by $\edit_\F(n)$. Let $F=|\F|$.

First, note that applying the generic approach for transitive metric spaces (as
with the Hamming space and the set difference space for small universe sizes)
does not work here, because the edit metric is not known to be transitive.
Instead, we consider embeddings of the edit metric on $\bit{n}$ into the
Hamming or set difference metric of much larger dimension. We look at two
types: standard low-distortion embeddings and ``biometric'' embeddings as
defined in \secref{bio-embed}.

For the binary edit distance space of dimension $n$, we obtain secure sketches and fuzzy extractors correcting $t$ errors with entropy loss roughly $tn^{o(1)}$, using a standard embedding, and $2.38 \sqrt[3]{tn\log n}$, using a relaxed embedding. The first technique works better when $t$ is small, say, $n^{1-\gamma}$ for a constant $\gamma>0$. The second technique is better when $t$ is large; it is meaningful roughly as long as $t<\frac{n}{15\log^2 n}$.

\subsection{Low-Distortion Embeddings}

A (standard) embedding with distortion $D$ is an injection $\psi:\M_1\into
\M_2$ such that for any two points $x,y\in\M_1$, the ratio
$\frac{\dis{\psi(x)}{\psi(y)}}{\dis{x}{y}}$ is at least 1 and at most $D$.

When the preliminary version of this paper appeared~\cite{DRS04}, no
nontrivial embeddings were known mapping edit distance into $\ell_1$
or the Hamming metric (i.e., known embeddings had distortion
$O(n)$). Recently, Ostrovsky and Rabani~\cite{OR05} gave an embedding
of the edit metric over $\F=\{0,1\}$ into $\ell_1$ with subpolynomial
distortion. It is an injective, polynomial-time computable embedding,
which can  be interpreted as mapping to the Hamming space $\bit{d}$,
where $d=\poly(n)$.~\footnote{The embedding of \cite{OR05} produces
strings of integers in the space $\set{1,\dots,O(\log n)}^{\poly(n)}$,
equipped with $\ell_1$ distance. One can convert this into the Hamming
metric with only a logarithmic blowup in length by representing each
integer in unary.}

\begin{fact}[\cite{OR05}]\label{fact:ORembed}
    There is a polynomial-time computable embedding \ifnum\siam=1
    denoted \fi
    $\psi_{\rm ed}:\edit_{\{0,1\}}(n)
    \into \bit{\poly(n)}$ with distortion $D_{\rm
    ed}(n)\defeq 2^{O(\sqrt{\log n\log\log n})}$.
\end{fact}

We can compose this embedding with the fuzzy extractor
constructions for the Hamming distance
to obtain a fuzzy extractor for edit distance
which will be good when $t$, the number of errors to be corrected, is quite
small.
 Recall that instantiating the syndrome fuzzy extractor construction
(\thmref{fe-upper}) with a BCH code allows one to correct $t'$ errors
out of $d$ at the cost of $t'\log d+2\logeps-2$ bits of entropy.

\begin{constr}
\label{constr:ORembed}
For any length $n$ and error threshold $t$, let $\psi_{\rm ed}$ be the
embedding given by \factref{ORembed} from $\edit_{\{0,1\}}(n)$ into $\bit{d}$ (where
$d=\poly(n)$), and let $\syn$ be the syndrome of a BCH code correcting $t' =
tD_{\rm ed}(n)$ errors in $\bit{d}$. Let
$\{H_x\}_{x\in X}$ be a family of universal hash functions from $\bit{d}$
to $\bit{\ell}$ for some $\ell$.
To compute $\gen$ on input $w\in \edit_{\{0,1\}}(n)$, pick a random $x$ and output
$$R=H_x(\psi_{\rm ed}(w))\ , P=(\syn(\psi_{\rm ed}(w)), x)\,.$$ 
To compute $\rep$ on inputs $w'$ and
$P=(s, x)$,  compute
$y=\rec(\psi_{\rm ed}(w'), s)$, where $\rec$ is from
\constrref{syndrome},
and output $R=H_x(y)$.
\end{constr}

Because $\psi_{\rm ed}$ is injective, a secure sketch can be
constructed similarly: $\ff(w)=\syn(\psi(w))$, and to recover $w$ from $w'$ and
$s$, compute $\psi_{\rm ed}^{-1}(\rec(\psi_{\rm ed}(w')))$.  However,
it is not known to be efficient, because it is not known how to
compute $\psi^{-1}_{\rm ed}$ efficiently.

\begin{prop}\label{prop:edit}
For any $n,t,m$, there is an average-case $(\edit_{\{0,1\}}(n), m, m',
\allowbreak t)$-secure sketch
and an efficient average-case $(\edit_{\{0,1\}}(n),m,\ell,t,\eps)$-fuzzy extractor 
where $m'=m-t 2^{O(\sqrt{\log n\log \log n})}$ and $\ell = m'-2\logeps+2$. 
In particular, for
any $\alpha<1$, there exists an efficient fuzzy extractor tolerating $n^\alpha$ errors
with entropy loss $n^{\alpha+o(1)}+2\logeps$.
\end{prop}

\begin{proof}
\ifnum\siam=0\sloppypar\fi
\constrref{ORembed} is the same as the construction
of \thmref{fe-upper} (instantiated with a BCH-code-based syndrome construction)
acting on $\psi_{\rm ed}(w)$.  Because $\psi_{\rm ed}$ is
injective, the min-entropy of $\psi_{\rm ed}(w)$ is the same
as the min-entropy $m$ of $w$.
The entropy loss in \constrref{syndrome} instantiated with BCH codes
is $t'\log d = t 2^{O(\sqrt{\log
n\log \log n})} \log \poly (n)$.  Because $2^{O(\sqrt{\log n\log \log n})}$
grows faster than $\log n$, this is the same as $t 2^{O(\sqrt{\log n\log
\log n})}$. 
\ifnum\siam=1\qquad\fi
\end{proof}

Note that the peculiar-looking distortion function from
\factref{ORembed} increases more slowly than any polynomial in $n$,
but still faster than any polynomial in $\log n$.  In sharp contrast,
the best lower bound states that any embedding of $\edit_{\zo}(n)$ into
$\ell_1$ (and hence Hamming) must have distortion at least
$\Omega(\log n / \log \log n)$ \cite{AK07}. Closing the gap between
the two bounds remains an open problem.

\mypar{General Alphabets} To extend the above construction to general
$\F$, we represent each character of $\F$ as a string of $\log F$
bits. This is an embedding $\F^n$ into $\bit{n\log F}$, which
increases edit distance by a factor of at most $\log F$.  Then $t'=t
(\log F) D_{\rm ed}(n)$ and $d=\poly(n, \log F)$. Using these
quantities, we get the generalization of \propref{edit} for
larger alphabets (again, by the same embedding) by changing the
formula for $m'$ to $m'= m-t(\log F) 2^{O(\sqrt{\log (n\log F) \log
\log (n \log F)})}$.

\subsection{Relaxed Embeddings for the Edit Metric}

In this section, we show that a relaxed notion of embedding, called a
\emph{biometric embedding} in \secref{bio-embed}, can produce fuzzy
extractors and secure sketches that are better than what one can get from
the embedding of \cite{OR05} when $t$ is large (they are also much simpler
algorithmically, which makes them more practical). We first discuss fuzzy
extractors and later extend the technique to secure sketches.

\mypar{Fuzzy Extractors}
Recall that unlike
low-distortion embeddings, biometric embeddings do not care about relative
distances, as long as points that were ``close'' (closer than
$t_1$) do not become ``distant'' (farther apart than $t_2$).  The only
additional requirement of a biometric embedding is that it preserve some
min-entropy: we do not want too many points to collide together.  We now describe such an embedding from the edit distance to the set difference.

A {\em $c$-shingle} is a length-$c$ consecutive substring
of a given string $w$. A {\em $c$-shingling} \cite{Bro97} of a string
$w$ of length $n$ is the set (ignoring order or repetition) of all
$(n-c+1)$ $c$-shingles of $w$.  (For instance, a 3-shingling of
``abcdecdeah'' is \{abc, bcd, cde, dec, ecd, dea, eah\}.) Thus, the
range of the $c$-shingling operation consists of all nonempty subsets
of size at most $n-c+1$ of $\F^c$.  Let
$\sdif(\F^c)$ stand for the set difference
metric over subsets of $\F^c$  and $\shin_c$ stand for the
$c$-shingling map from $\edit_\F(n)$ to $\sdif(\F^c)$.
We now show that
$\shin_c$ is a good biometric embedding.

\begin{lemma}\label{lem:edit}
For any $c$,  $\shin_c$ is an average-case $(t_1, t_2 = (2c-1)t_1, m_1, m_2 = m_1 -
\lceil \frac{n}{c} \rceil
\log_2 (n-c+1))$-biometric embedding of $\edit_\F(n)$ into
$\sdif(\F^c)$.
\end{lemma}

\begin{proof}
Let $w, w'\in \edit_\F(n)$ be such that $\dis{w}{w'} \le t_1$ and $I$ be
the sequence
of at most $t_1$ insertions and deletions that transforms $w$ into $w'$. It
is easy to see that each character deletion or insertion
adds at most $(2c-1)$ to the symmetric difference between
$\shin_c(w)$ and $\shin_c(w')$, which implies that
$\dis{\shin_c(w)}{\shin_c(w')} \le (2c-1)t_1$, as needed.

For $w\in \F^n$, define $g_c(w)$ as follows.  Compute
$\shin_c(w)$ and store the resulting shingles in lexicographic order
$h_1\ldots h_k$ ($k \leq n-c+1$). Next, naturally partition $w$ into
$\lceil n/c\rceil$ $c$-shingles $s_1\ldots s_{\lceil n/c \rceil}$, all
disjoint except for (possibly) the last two, which overlap by $c\lceil
n/c \rceil-n$ characters. Next, for $1\le j \le \lceil n/c \rceil$,
set $p_j$ to be the index $i\in \set{0\ldots k}$ such that $s_j =
h_i$. In other words, $p_j$ tells the index of the $j$th disjoint
shingle of $w$ in the alphabetically ordered $k$-set $\shin_c(w)$.
Set $g_c(w) = (p_1, \dots, p_{\lceil n/c\rceil})$.  (For instance,
$g_3($``abcdecdeah''$)=(1, 5, 4, 6)$, representing the alphabetical
order of ``abc'', ``dec'', ``dea'' and ``eah'' in
$\shin_3($``abcdecdeah''$)$.)  The number of possible values for
$g_c(w)$ is at most $(n-c+1)^{\lceil\frac{n}{c}\rceil}$, and $w$
can be completely recovered from $\shin_c(w)$ and $g_c(w)$.

Now, assume $W$ is any distribution of min-entropy at least $m_1$ on
$\edit_\F(n)$. Applying~\lemref{hinf}(b), we get $\thinf(W\mid g_c(W)) \ge m_1\
- \lceil \frac{n}{c} \rceil
 \log_2 (n-c+1)$.  Since $\Pr(W=w\mid g_c(W) = g) =
\Pr(\shin_c(W)=\shin _c(w) \mid g_c(W) = g)$ (because given $g_c(w)$,
$\shin_c(w)$ uniquely determines $w$ and vice versa), by applying
the definition of $\thinf$, we obtain
$\hinf(\shin_c(W)) \ge \thinf(\shin_c(W) \mid g_c(W)) =
\thinf(W\mid g_c(W))$.  The same proof
holds for average min-entropy, conditioned on some auxiliary information $I$.
\ifnum\siam=1\qquad\fi
\end{proof}

By \thmref{bch}, for universe $\F^c$ of size $F^c$ and distance
threshold $t_2 = (2c-1)t_1$, we can construct a secure sketch for the
set difference metric with entropy loss $t_2 \lceil \log (F^c+1) \rceil
$ ($\lceil \cdot \rceil$ because \thmref{bch} requires the universe
size to be one less than a power of 2).  By~\lemref{ff-2wise}, we can obtain
a fuzzy extractor from such a sketch, with additional entropy loss
$2\logeps-2$.  Applying~\lemref{embedfe} to the above embedding and
this fuzzy extractor, we obtain a fuzzy extractor for $\edit_\F(n)$, any
input entropy $m$, any distance $t$, and any security parameter
$\eps$, with the following entropy loss:
\[
\left \lceil \frac{n}{c} \right \rceil \cdot
\log_2 (n-c+1)+(2c-1)t\lceil \log (F^c+1)\rceil+2\logeps-2\,
\]
(the first component of the entropy loss comes from the embedding, the
second from the secure sketch for set difference, and the third from
the extractor).  The above sequence of lemmas results in the following
construction, parameterized by shingle length $c$ and a family of
universal hash functions $\Hfam=\{\sdif(\F^c) \to \zo^l\}_{x\in
X}$, where $l$ is equal to the input entropy $m$ minus the entropy
loss above.

\begin{constr}[Fuzzy Extractor for Edit Distance]
\label{constr:edit-extractor} \ifnum\siam=0 \ \\ \fi
To compute $\gen(w)$ for $|w|=n$:
\ifnum\siam=0
\begin{CompactEnumerate}
\else
\begin{remunerate}
\fi
\item Compute $\shin_c(w)$ by computing
$n-c+1$ shingles $(v_1, v_2, \dots, v_{n-c+1})$ and
 removing duplicates to form the shingle set $v$ from $w$.
\item Compute $s=\syn(x_v)$ as in \constrref{bch}.
\item Select a hash function $H_x\in \Hfam$ and output
  $(R=H_x(v), P=(s, x))$.
\ifnum\siam=0
\end{CompactEnumerate}
\else
\end{remunerate}
\fi
To compute $\rep(w', (s, x))$:
\ifnum\siam=0
\begin{CompactEnumerate}
\else
\begin{remunerate}
\fi
\item Compute $\shin_c(w')$ as above to get $v'$.
\item Use $\rec(v', s)$ from in \constrref{bch} to recover $v$.
\item Output $R=H_x(v)$.
\ifnum\siam=0
\end{CompactEnumerate}
\else
\end{remunerate}
\fi
\end{constr}

\ifnum\siam=0\noindent\fi
We thus obtain the following theorem.

\begin{theorem}
For any $n, m, c$ and $0<\eps\le 1$, there is an efficient average-case
$(\edit_\F(n), m, m-\lceil \frac{n}{c} \rceil
\log_2 (n-c+1)-(2c-1)t\lceil \log (F^c+1) \rceil-2\logeps+2, t, \eps)$-fuzzy
extractor.
\end{theorem}

Note that the choice of $c$ is a parameter; by ignoring $\lceil \cdot
\rceil$ and replacing $n-c+1$ with $n$, $2c-1$ with $2c$ and $F^c+1$
with $F^c$, we get that the
minimum entropy loss occurs near
\[
c=\left(\frac{n\log n}{4t\log F}\right)^{1/3}
\]
and is about $2.38\left(t\log F\right)^{1/3}\left( n \log n\right)^{2/3}$
($2.38$ is really $\sqrt[3]{4}+1/\sqrt[3]{2}$).  In particular, if the
original string has a linear amount of entropy $\theta(n\log F)$, then we
can tolerate $t=\Omega(n\log^2 F /\log^2 n)$ insertions and deletions while
extracting $\theta(n\log F)-2\logeps$ bits.  The number of bits
extracted is linear; if the string length $n$ is polynomial in
the alphabet size $F$, then the number of errors tolerated is linear also.

\mypar{Secure Sketches}
Observe that the proof of~\lemref{edit} actually demonstrates that our
biometric embedding based on shingling is an embedding with recovery
information $g_c$.  Observe also that it is easy to reconstruct $w$ from
$\shin_c(w)$ and $g_c(w)$.  Finally, note that PinSketch (\constrref{bch})
is an average-case secure sketch (as are
all secure sketches in this work).  Thus, combining~\thmref{bch}
with~\lemref{embedff}, we obtain the following theorem.

\begin{constr}[Secure Sketch for Edit Distance]
For $\ff(w)$, compute $v=\shin_c(w)$ and $s_1=\syn(x_v)$ as in
\constrref{edit-extractor}.   Compute $s_2=g_c(w)$,
writing each $p_j$ as a string of $\lceil \log n \rceil$ bits.  Output
$s=(s_1, s_2)$. For $\rec(w', (s_1, s_2))$,  recover $v$ as in
\constrref{edit-extractor},  sort it in alphabetical order, and
recover $w$ by stringing along elements of $v$ according to indices in
$s_2$.
\end{constr}

\begin{theorem}
For any $n, m, c$ and $0<\eps\le 1$, there is an efficient average-case 
$(\edit_\F(n), m, m-\lceil \frac{n}{c} \rceil
\log_2 (n-c+1)-(2c-1)t\lceil \log (F^c+1)\rceil, t)$
secure sketch.
\end{theorem}

\ifnum\siam=0\noindent\fi
The discussion about optimal values of $c$ from above applies equally here.

\ifnum\siam=1
\textit{Remark.\ }
\else
\begin{remark}
\fi
In our definitions of secure sketches and fuzzy extractors, we required the
original $w$ and the (potentially) modified $w'$ to come from the same
space $\M$.  This requirement was for simplicity of exposition.  We can
allow $w'$ to come from a larger set, as long as distance from $w$ is
well-defined.  In the case of edit distance, for instance, $w'$ can be
shorter or longer than $w$; all the above results will apply as long
as it is still within $t$ insertions and deletions.
\ifnum\siam=0
\end{remark}
\fi

%%%%%%%%%%%%%%%%%%%%%%%%%%%%%%%%%%%%%%%%%%%%%%%%%%%%%%%%%%%%%%%%

\section{Probabilistic Notions of Correctness}
\label{sec:improved-param-list}

The error model considered so far in this work is very strong: we
required that secure sketches and fuzzy extractors accept \emph{every}
secret $w'$ within distance $\erc$ of the original input $w$, with no probability of error.

Such a stringent model is useful as it makes no assumptions on either the exact stochastic properties of the error process or the adversary's computational limits. However, \lemref{ff-lower} shows that secure sketches (and fuzzy extractors) correcting $\erc$ errors can only be as ``good'' as error-correcting codes with minimum distance $2\erc+1$. By slightly relaxing the correctness condition, we will see that one can tolerate many more errors.
For example, there is no good code which can correct $n/4$ errors in the
binary Hamming metric: by the Plotkin bound (see, e.g., \cite[Lecture 8]{Sud01}) a code with minimum distance greater than $n/2$ has at most $2n$ codewords. Thus, there is no secure sketch with residual entropy $m'\geq \log n$ which can
correct $n/4$ errors with probability 1. However, with the relaxed
notions of correctness below, one can tolerate arbitrarily close to
$n/2$ errors, i.e., correct
$n(\frac 1 2 -\gamma)$ errors for any constant $\gamma>0$, and still have residual entropy $\Omega(n)$.

\newcommand{\err}{\alpha}
\newcommand{\logerr}{\log\paren{\frac 1 \alpha}}

In this section, we discuss three relaxed error models and
show how the constructions of the previous sections can be modified to
gain greater error-correction in these models. We will focus on secure sketches for the binary Hamming metric. The same constructions yield fuzzy extractors  (by \lemref{ff}). Many of the observations here also apply to metrics other than Hamming.

A common point is that we will require only that the a corrupted input $w'$ be recovered with probability at least $1-\err<1$ (the probability space varies). 
We describe each model in terms of the additional assumptions made on the error process. We describe constructions for each model in the subsequent sections.

\ifnum\siam=0
\begin{description}
\fi

\ifnum\siam=0
\item[Random Errors.]
\else
\paragraph{Random Errors (\secref{random-errors})}
\fi
 Assume there is a \emph{known} distribution on the errors which occur in the data. For the Hamming metric, the most common distribution is the binary symmetric channel $BSC_p$: each bit of the input is flipped with probability $p$ and left untouched with probability $1-p$. We require that for any input $w$, $\rec(W',\ff(w))=w$ with probability at least $1-\err$ over the coins of $\ff$ and over $W'$ drawn applying the noise distribution to $w$.

In that case, one can correct an error rate up to Shannon's bound on noisy channel coding. This bound is tight.
Unfortunately, the assumption of a known noise process is too strong for most applications: there is no reason to believe we understand the exact distribution on errors which occur in complex data such as biometrics.%
\footnote{Since the assumption here plays a role only in correctness, it is
still more reasonable than assuming that we know exact distributions on the data in proofs of \emph{secrecy}.  However, in both cases, we would like to enlarge the class of distributions for which we can provably satisfy the definition of security.}
However, it provides a useful baseline by which to measure results for other models.

\ifnum\siam=0
\item[Input-dependent Errors.]
\else
\paragraph{Input-dependent Errors (\secref{input-dep})}
\fi
 The errors are adversarial, subject only
  to the conditions that (a) the error magnitude $\dis{w}{w'}$ is bounded to a
  maximum of $\erc$, and (b) the corrupted word
  \emph{depends only on the input $w$}, and  not on the secure sketch
  $\ff(w)$. Here we require that for any pair $w,w'$ at distance at
  most $\erc$, we have $\rec(w',\ff(w))=w$ with probability at least
  $1-\err$ over the coins of  $\ff$.

This model encompasses any complex noise process which has been observed to never introduce more than $\erc$ errors. Unlike the assumption of a particular distribution on the noise, the bound on magnitude can be checked experimentally. Perhaps surprisingly, in this model we can tolerate just as large an error rate as in the model of random errors. That is, we can tolerate an error rate up to Shannon's coding bound and no more.

\ifnum\siam=0
\item[Computationally bounded Errors.]
\else
\paragraph{Computationally bounded Errors (\secref{list-decoding})}
\fi
 The errors are adversarial and may depend on both $w$ and the publicly stored information $\ff(w)$. However, we assume that the errors are introduced by a process of bounded computational power. That is, there is a probabilistic circuit of polynomial size (in the length $n$) which computes $w'$ from $w$. The adversary cannot, for example, forge a digital signature and base the error pattern on the signature.

It is not clear whether this model allows correcting errors up to the Shannon
bound, as in the two models above. The question is related to open questions on
the construction of efficiently list-decodable codes. However, when the error
rate is either very high or very low, then the appropriate list-decodable codes
exist and we can indeed match the Shannon bound.
\ifnum\siam=0
\end{description}
\fi

\mypar{Analogues for Noisy Channels and the Hamming Metric}
\label{sec:hamming-metric}
Models analogous to the ones above have been studied in the
literature on codes for noisy binary channels (with the Hamming
metric). Random errors and computationally bounded errors both make
obvious sense in the coding context~\cite{Sha48,MPSW05}. The second
model --- input-dependent errors --- does not immediately make sense
in a coding situation, since there is no data other than the
transmitted codeword on which errors could depend. Nonetheless,
there is a natural, analogous model for noisy channels: one can
allow the sender and receiver to share either (1) common, secret
random coins (see \cite{DGL04,Lan04} and references therein) or (2)
a side channel with which they can communicate a small number of
noise-free, secret bits~\cite{Gur03}.

Existing results on these three models for the Hamming metric can be
transported to our context using the code-offset
construction:
$$
\ff(w;x) = w \oplus C(x)\,.$$
Roughly, any code which corrects
errors in the models above will lead to a secure sketch (resp. fuzzy
extractor) which corrects errors in the model.
We explore the consequences for each of the three models in the next sections.

\subsection{Random Errors}
\label{sec:random-errors}

The random error model was famously considered by
Shannon~\cite{Sha48}. He showed that for any discrete, memoryless
channel, the rate at which information can be reliably transmitted is
characterized by the maximum mutual information between the inputs and
outputs of the channel. For the binary symmetric channel with
crossover probability $p$, this means that there exist codes encoding
$k$ bits into $n$ bits, tolerating error probability $p$ in each bit
if and only if
$$\frac{k}{n} < 1-h(p) - \delta(n)\,,$$ where $h(p)=-p\log p
-(1-p)\log(1-p)$ and $\delta(n)=o(1)$.  Computationally efficient
codes achieving this bound were found later, most notably by
Forney~\cite{For66}. We can use the code-offset construction $\ff(w;x)
= w \oplus C(x)$ with an appropriate concatenated code~\cite{For66}
or, equivalently, $\ff(w) = \syn_C(w)$ since the codes can be
linear. We obtain:

\begin{prop}\label{prop:randerr}
  For any error rate $0<p<1/2$ and constant $\delta>0$, for large
  enough $n$ there exist secure sketches with entropy loss
  $(h(p)+\delta)n$, which correct the error rate of $p$ in the data with
  high probability (roughly $2^{-c_\delta n}$ for a constant $c_\delta>0$).

  The probability here is taken over the \emph{errors} only (the distribution on input strings $w$ can be arbitrary).
\end{prop}

The quantity $h(p)$ is less than 1 for any $p$ in the range $(0,1/2)$.
In particular, one can get nontrivial secure sketches even for a very
high error rate $p$ as long as it is less than $1/2$; in contrast, no secure sketch which corrects errors with probability 1 can tolerate $t\geq n/4$.  Note that several other works on biometric
cryptosystems consider the model of randomized errors and obtain
similar results, though the analyses assume that the distribution on inputs is uniform~\cite{TG04,CZ04}.

\mypar{A Matching Impossibility Result}
The bound above is tight. The matching impossibility result also applies to
input-dependent and computationally bounded errors, since random errors are a special case of both more complex models.

We start with an intuitive argument: If a secure sketch allows recovering from random errors with high probability, then it must contain enough information about $w$ to describe the error pattern (since given $w'$ and $\ff(w)$, one can recover the error pattern with high probability). Describing the outcome of $n$ independent coin flips with probability $p$ of heads requires $nh(p)$ bits, and so the sketch must reveal $nh(p)$ bits about $w$.

In fact, that argument simply shows that $nh(p)$ bits of Shannon information are leaked about $w$, whereas we are concerned with min-entropy loss as defined in \secref{defs}. To make the argument more formal, let $W$ be uniform over $\bit{n}$ and observe that with high probability over the output of the sketching algorithm, $v=\ff(w)$, the conditional distribution $W_v = W|_{\ff(W)=v}$ forms a good code for the binary symmetric channel. That is, for most values $v$, if we sample a random string $w$ from $W|_{\ff(W)=v}$ and send it through a binary symmetric channel, we will be able to recover the correct value $w$.
That means there exists some $v$ such that both (a) $W_v$ is a good code and (b) $\hinf(W_v)$ is close to $\thinf(W|\ff(W))$. Shannon's noisy coding theorem says that such a code can have entropy at most $n(1-h(p)+o(1))$. Thus the construction above is optimal:

\begin{prop}\label{prop:randlower}
  For any error rate $0<p<1/2$, any secure sketch $\ff$ which corrects random errors (with rate $p$) with probability at least $2/3$ has entropy loss at least $n(h(p)-o(1))$; that is, $\thinf(W|\ff(W))\leq n(1-h(p)-o(1))$ when $W$ is drawn uniformly from $\bit{n}$.
\end{prop}

\subsection{Randomizing Input-dependent Errors}
\label{sec:input-dep}
Assuming errors distributed randomly according to a known
distribution seems very limiting. In the Hamming metric, one can
construct a secure sketch which achieves the same result as with
random errors for every error process where the magnitude of the
error is bounded, as long as the errors are independent of the
output of $\ff(W)$. The same technique was used previously by
Bennett et al. \cite[p. 216]{BBR88} and, in a slightly different
context, Lipton \cite{Lip94,DGL04}.

The idea is to choose a random permutation $\pi:[n]\to [n]$, permute the bits of $w$ before applying the sketch, and store the permutation $\pi$ along with $\ff(\pi(w))$. Specifically, let $C$ be a linear code tolerating a $p$ fraction of random errors with redundancy $n-k\approx nh(p)$. Let
$$\ff(w;\pi) = (\pi,\ \syn_C(\pi(w)))\,,$$ where $\pi:[n]\to[n]$ is a
random permutation and, for $w=w_1\cdots w_n\in \bit{n}$,  $\pi(w)$ denotes
the permuted string $w_{\pi(1)}w_{\pi(2)}\cdots w_{\pi(n)}$. The recovery
algorithm operates in the obvious way: it first permutes the input $w'$
according to $\pi$ and then runs the usual syndrome recovery algorithm to recover $\pi(w)$.

For any particular pair $w,w'$, the difference $w\oplus w'$ will be mapped to a random vector of the same weight by $\pi$, and any code for the binary symmetric channel (with rate $p \approx t/n$) will correct such an error with high probability.

Thus we can construct a sketch with entropy loss $n (h(t/n)-o(1))$ which corrects any $t$ flipped bits with high probability. This is optimal by the lower bound for random errors (\propref{randlower}), since a sketch for data-dependent errors will also correct random errors. It is also possible to reduce the amount of randomness, so that the {\em size} of the sketch meets the same optimal bound \cite{Smi07}.

An alternative approach to input-dependent errors is discussed in the
last paragraph of \secref{list-decoding}.

\subsection{Handling Computationally Bounded Errors Via List Decoding}
\label{sec:list-decoding}

As mentioned above, many results on noisy coding for other error models in
Hamming space extend to secure sketches.
The previous sections discussed random, and randomized, errors. In this section, we discuss
constructions \cite{Gur03,Lan04,MPSW05} which transform a \emph{list-decodable} code, defined below, into uniquely decodable codes for a
particular error model.
These transformations can also be used in the setting of secure sketches, leading to better tolerance of computationally bounded errors. For some ranges of parameters, this yields optimal sketches, that is, sketches which meet the Shannon bound on the fraction of tolerated errors.

\mypar{List-Decodable Codes}
A code $C$ in a metric space $\M$ is called \emph{list-decodable} with
list size $L$ and distance $\erc$ if for every point $x\in\M$, there
are at most $L$ codewords within distance $\erc$ of $\M$. A
list-decoding algorithm takes as input a word $x$ and returns the
corresponding list $c_1,c_2,\dots$ of codewords.  The most interesting
setting is when $L$ is a small polynomial (in the description size
$\log|\M|$), and there exists an efficient list-decoding algorithm. It
is then feasible for an algorithm to go over each word in the list and
accept if it has some desirable property. There are many examples of
such codes for the Hamming space; for a survey see Guruswami's
thesis~\cite{Gur01}. Recently there has been significant progress in constructing list-decodable codes for large alphabets, e.g., \cite{PV05,GR06}. 

Similarly, we can define a \emph{list-decodable secure sketch} with
size $L$ and distance $\erc$ as follows: for any pair of words $w,
w'\in \M$ at distance at most $\erc$, the algorithm $\Rec(w',\ff(w))$
returns a list of at most $L$ points in $\M$; if $\disfn(w,w')\leq
\erc$, then one of the words in the list must be $w$ itself. The
simplest way to obtain a list-decodable secure sketch is to use the
code-offset construction of \secref{hamming} with a list-decodable
code for the Hamming space. One obtains a different example by running
the improved Juels-Sudan scheme for set difference (\constrref{ijs}),
replacing ordinary decoding of Reed-Solomon codes with list
decoding. This yields a significant improvement in the number of
errors tolerated at the price of returning a list of possible
candidates for the original secret.

\mypar{Sieving the List}
\label{sec:sieving-list}
Given a list-decodable secure sketch $\ff$, all that's needed is to
store some additional information which allows the receiver to
disambiguate $w$ from the list. Let's suggestively name the additional
information $Tag(w;R)$, where $R$ is some additional randomness
(perhaps a key). Given a list-decodable code $C$, the sketch will
typically look like
$$\ff(w;x )=(\ w \oplus C(x) , \ Tag(w)\ )\,.$$
On inputs $w'$ and $(\Delta, tag)$, the recovery algorithm consists of
running the list-decoding algorithm on $w'\oplus \Delta$ to obtain a
list of possible codewords $C(x_1),\dots,C(x_L)$. There is a
corresponding list of candidate inputs $w_1,\dots,w_L$, where $w_i =
C(x_i) \oplus \Delta$, and the algorithm outputs the first $w_i$ in
the list such that $Tag(w_i) = tag$. We will choose the function
$Tag()$ so that the adversary can not arrange to have two values in
the list with valid tags.

We consider two $Tag()$ functions, inspired by
\cite{Gur03,Lan04,MPSW05}.

\newcommand{\key}{\mathit{key}}

\ifnum\siam=0
\begin{enumerate}
\fi
\ifnum\siam=0
\item 
\else
1. 
\fi
Recall that for computationally bounded errors, the corrupted string  $w'$ depends on \emph{both} $w$ and
  $\ff(w)$, but $w'$ is computed by a probabilistic circuit of size polynomial in $n$.

  Consider  $Tag(w) = \hash(w)$, where $\hash$ is drawn from a
  collision-resistant function family. More specifically, we will use some extra randomness $r$ to choose a key $\key$ for a collision-resistant hash family. The output of the sketch is then
$$\ff(w; x,r) = (\ w \oplus C(x) ,\ \key(r),\ \hash_{\key(r)}(w)\  ).$$
If the list-decoding algorithm for the code $C$ runs in polynomial time, then the  adversary succeeds only if
  he can find a value $w_i\neq w$ such that
$\hash_{\key}(w_i)=\hash_{\key}(w)$, that is, only by finding a collision
for the hash function. By assumption, a polynomially bounded adversary succeeds only with negligible probability.

    The additional entropy loss, beyond that of the code-offset part of the sketch, is bounded above by the output
  length of the hash function. If $\err$ is the
  desired bound on the adversary's success probability, then for standard assumptions on hash functions this loss will be polynomial in $\log(1/\err)$.

  In principle this transformation can yield sketches which achieve the optimal entropy loss $n(h(t/n)-o(1))$, since codes with polynomial list size $L$ are known to exist for error rates approaching the Shannon bound. However, in order to use the construction the code must also be equipped with a reasonably efficient algorithm for finding such a list. This is necessary both so that recovery will be efficient and, more subtly, for the proof of security to go through (that way we can assume that the polynomial-time adversary  knows the list of words generated during the recovery procedure).
  We do not know of {\em efficient} (i.e., polynomial-time
  constructible and decodable) binary list-decodable codes which meet
  the Shannon bound for all choices of parameters.  However, when the
  error rate is near $\half$ such codes are known~\cite{GS00}. Thus,
  this type of construction yields essentially optimal sketches when
  the error rate is near $1/2$. This is quite similar to analogous
  results on channel coding~\cite{MPSW05}.  Relatively little is known
  about the performance of efficiently list-decodable codes in other
  parameter ranges for binary alphabets~\cite{Gur01}.

\ifnum\siam=0
\item
\else
2.
\fi 
A similar, even simpler, transformation can be used in the
  setting of input-dependent errors (i.e., when the errors depend only on the
  input and not on the sketch, but the adversary is not assumed to be
  computationally bounded).
One can store $Tag(w) = (I,h_I(w))$, where
  $\set{h_i}_{i\in \I}$ comes from a universal hash family mapping from
  $\M$ to $\bit{\ell}$, where $\ell = \logerr + \log L$ and $\err$ is
  the probability of an incorrect decoding.

  The proof is simple: the values $w_1,\dots,w_L$ do not depend on $I$,
  and so for any value $w_i\neq w$, the probability that
  $h_I(w_i)=h_I(w)$ is $2^{-\ell}$. There are at most $L$ possible
  candidates, and so the probability that any one of the elements in
  the list is accepted is at most $L\cdot 2^{-\ell}=\err$
  The additional entropy loss incurred is at most
  $\ell=\logerr+\log(L)$.

  In principle, this transformation can do as well as the randomization approach of the previous section. However, we do not know of efficient binary list-decodable codes meeting the Shannon bound for most parameter ranges.  Thus, in general, randomizing the errors (as in the previous section) works better in the input-dependent setting.
\ifnum\siam=0
\end{enumerate}
\fi

%%%%%%%%%%%%%%%%%%%%%%%%%%%%%%%%%%%%%%%%%%%%%%%%%%%%%%%%%%%%%%%%

\section{Secure Sketches and Efficient Information Reconciliation}
\label{sec:reconciliation}
Suppose Alice holds a set $w$ and Bob holds a set $w'$ that are
close to each other.  They
wish to reconcile the sets: to discover the symmetric difference
$w \triangle w'$ so that they can take whatever appropriate
(application-dependent) action to make their two sets agree.
Moreover, they wish to do this communication-efficiently, without
having to transmit entire sets to each other.  This problem is known as
set reconciliation and naturally arises in various settings.

Let $(\ff, \rec)$ be a secure sketch for set difference that can
handle distance up to $t$; furthermore, suppose that $|w\triangle w'|
\le t$.  Then if Bob receives $s=\ff(w)$ from Alice, he will be able
to recover $w$, and therefore $w \triangle w'$, from $s$ and $w'$.
Similarly, Alice will be able find $w \triangle w'$ upon receiving
$s'=\ff(w')$ from Bob.  This will be communication-efficient if $|s|$
is small.  Note that our secure sketches for set difference of
Sections~\ref{sec:ijs} and ~\ref{sec:sublin} are indeed short---in fact, they
are secure precisely because they are short.  Thus, they also make
good set reconciliation schemes.

Conversely, a good (single-message) set reconciliation scheme makes a
good secure sketch: simply make the message the sketch.  The entropy
loss will be at most the length of the message, which is short in a
communication-efficient scheme.  Thus, the set reconciliation scheme
CPISync of~\cite{MTZ03} makes a good secure sketch.  In fact, it is
quite similar to the secure sketch of \secref{ijs}, except instead of
the top $t$ coefficients of the characteristic polynomial it uses the
values of the polynomial at $t$ points.

PinSketch of \secref{sublin}, when used for set
reconciliation, achieves the same parameters as CPISync
of~\cite{MTZ03}, except decoding is faster, because instead of
spending $t^3$ time to solve a system of linear equations, it spends
$t^2$ time for Euclid's algorithm.  Thus, it can be substituted
wherever CPISync is used, such as PDA synchronization~\cite{STA03} and
PGP key server updates~\cite{Minsky-SKS}.  Furthermore, optimizations
that improve computational complexity of CPISync through the use of
interaction~\cite{MT02} can also be applied to PinSketch.

Of course, secure sketches for other metrics are similarly related to
information reconciliation for those metrics.  In particular, ideas for
edit distance very similar to ours were independently considered in
the context of information reconciliation by~\cite{CT04}.

\ifnum\siam=0
\section*{Acknowledgments}

This work evolved over several years and discussions with many people
enriched our understanding of the material at hand. In roughly
chronological order, we thank Piotr Indyk for discussions about embeddings
and for his help in the proof of~\lemref{edit}; Madhu Sudan, for helpful
discussions about the construction of \cite{JS02} and the uses of
error-correcting codes; Venkat Guruswami, for enlightenment about list
decoding; Pim Tuyls, for pointing out relevant previous work; Chris
Peikert, for pointing out the model of computationally bounded adversaries
from \cite{MPSW05}; Ari Trachtenberg, for finding an error in the
preliminary version of \appref{bch}; Ronny Roth, for discussions about
efficient BCH decoding; Kevin Harmon and Soren Johnson, for their
implementation work; and Silvio Micali and anonymous referees,
for suggestions on presenting our
results.

The work of the Y.D. was partly funded by the National Science
Foundation under CAREER Award No. CCR-0133806 and Trusted Computing
Grant No. CCR-0311095, and by the New York University Research
Challenge Fund 25-74100-N5237.  The work of the L.R. was partly funded
by the National Science Foundation under Grant Nos. CCR-0311485, CCF-0515100 and CNS-0202067.  The
work of the A.S. at MIT was partly funded by US A.R.O. grant
DAAD19-00-1-0177 and by a Microsoft Fellowship. While at the Weizmann Institute,
A.S. was supported by the Louis L. and Anita M. Perlman Postdoctoral
Fellowship.

%Following line adds a ``References'' to the table of contents

\addcontentsline{toc}{section}{References}
%Included for Gather Purpose only:
%GATHER{fuzzy.bib}
\bibliographystyle{alpha}
\bibliography{fuzzy}
\fi

\appendix
%\newpage
%

\section{Proof of \lemref{hinf}}\label{sec:min-prop}

Recall that \lemref{hinf} considered random variables $A,B,C$ and
consisted of two parts, which we prove one after the other.

Part (a) stated that for any $\delta>0$, the conditional entropy
 $\hinf(A|B=b)$ is at least $\thinf(A|B)-\log(1/\delta)$ with
 probability at least $1-\delta$ (the probability here is taken over
 the choice of $b$).
Let $p=2^{-\thinf(A\mid B)}=\expe{b}{2^{-\hinf(A\mid B=b)}}$. By the
Markov inequality, $2^{-\hinf(A\mid B=b)} \leq p/\delta$ with
probability at least $1-\delta$. Taking logarithms, part (a) follows.

Part (b) stated that if $B$ has at most $2^\loss$ possible values,
then $\thinf(A\mid (B, C)) \geq \thinf((A,B)\mid C) - \loss \geq
\thinf(A\mid C)-\loss$. In particular, $\thinf(A\mid B) \geq \hinf((A,
B))-\loss \geq \hinf(A) - \loss$. Clearly, it suffices to prove the
first assertion (the second follows from taking $C$ to be
constant). Moreover, the second inequality of the first assertion
follows from the fact that $\Pr[A=a \wedge B=b \mid C=c]\le
\Pr[A=a\mid C=c]$, for any $c$. Thus, we prove only that $\thinf(A\mid
(B, C)) \geq \thinf((A,B)\mid C) - \loss$:

\ifnum\siam=0
\begin{eqnarray*}
\thinf(A\mid(B, C)) &  =& -\log \expe{(b, c)\from (B, C)}{\max_a
\Pr[A=a\mid B=b
\wedge C=c]} \\
& = & -\log \sum_{(b, c)} \max_a \Pr[A=a \mid B=b \wedge C=c]\Pr[B=b\wedge C=c]\\
& = & -\log \sum_{(b, c)} \max_a \Pr[A=a \wedge B=b \mid C=c] \Pr[C=c]\\
& = & -\log \sum_b \expe{c \from C}{ \max_a \Pr[A=a \wedge B=b \mid C=c]}  \\
& \geq & -\log \sum_{b} \expe{c \from C}{ \max_{a,b'} \Pr[A=a \wedge B=b' \mid C=c]}  \\
& = & -\log \sum_b 2^{-\thinf((A, B)\mid C)} \ge -\log 2^\loss
2^{-\thinf((A, B)\mid C)}
=
\thinf((A, B)\mid
C)-\loss\,.
\end{eqnarray*}
\else
\begin{eqnarray*}
\thinf(A\mid(B, C)) &  =& -\log \expe{(b, c)\from (B, C)}{\max_a
\Pr[A=a\mid B=b
\wedge C=c]} \\
& = & -\log \sum_{(b, c)} \max_a \Pr[A=a \mid B=b \wedge C=c]\Pr[B=b\wedge C=c]\\
& = & -\log \sum_{(b, c)} \max_a \Pr[A=a \wedge B=b \mid C=c] \Pr[C=c]\\
& = & -\log \sum_b \expe{c \from C}{ \max_a \Pr[A=a \wedge B=b \mid C=c]}  \\
& \geq & -\log \sum_{b} \expe{c \from C}{ \max_{a,b'} \Pr[A=a \wedge B=b' \mid C=c]}  \\
& = & -\log \sum_b 2^{-\thinf((A, B)\mid C)} \ge -\log 2^\loss
2^{-\thinf((A, B)\mid C)}
\\ & = &
\thinf((A, B)\mid
C)-\loss\,.
\end{eqnarray*}
\fi
The first inequality in the above derivation holds since taking the
maximum over all pairs $(a,b')$ (instead of over pairs $(a,b)$ where
$b$ is fixed) increases the terms of the sum and hence decreases
the negative log of the sum.

\section{On Smooth Variants of Average Min-Entropy and the Relationship
  to Smooth R\'enyi Entropy}
\label{sec:smooth}

Min-entropy is a rather fragile measure: a single high-probability
element can ruin the min-entropy of an otherwise good distribution.
This is often circumvented within proofs by considering a distribution which is close to the distribution of interest, but which has higher entropy.
Renner and Wolf~\cite{RW04} systematized this approach with
the notion of \emph{$\eps$-smooth} min-entropy (they use the term
``R\'enyi entropy of order $\infty$'' instead of ``min-entropy''),
which considers all distributions that are $\eps$-close:
\[
\hepsinf(A) = \max_{B:\ \sd{A,B}\le \eps} \hinf (B)\,.
\]
Smooth min-entropy very closely relates to the amount of extractable
nearly uniform randomness: if one can map $A$ to a distribution that
is $\eps$-close to $U_m$, then $\hepsinf(A)\ge m$; conversely,
from any $A$ such that $\hepsinf(A)\ge m$, and for any $\eps_2$,
one can extract $m-2\logepstwo$ bits that are
$\eps+\eps_2$-close to uniform (see~\cite{RW04} for a more precise
statement; the proof of the first statement follows by considering the
inverse map, and the proof of the second from the leftover hash lemma,
which is discussed in more detail in \lemref{2wise}).  For some
distributions, considering the smooth min-entropy will improve the
number and quality of extractable random bits.

A smooth version of average min-entropy can also be considered,
defined as
\[
\thepsinf(A\mid B) = \allowbreak \max_{(C, D):\ \sd{(A,B), (C, D)}\le
\eps}\thinf(C\mid D)\,.
\]
It similarly relates very closely to the number of extractable bits
that look nearly uniform to the adversary who knows the value of $B$,
and is therefore perhaps a better measure for the quality of a secure
sketch that is used to obtain a fuzzy extractor.
All our results can be cast in terms of smooth entropies throughout,
with appropriate modifications (if input entropy is $\eps$-smooth,
then output entropy will also be $\eps$-smooth, and extracted random
strings will be $\eps$ further away from uniform).  We avoid doing so
for simplicity of exposition.  However, for some input distributions,
particularly ones with few elements of relatively high probability,
this will improve the result by giving more secure sketches or
longer-output fuzzy extractors.

Finally, a word is in order on the relation of average min-entropy to
conditional min-entropy, introduced by Renner and Wolf in~\cite{RW05},
and defined as $\hinf(A\mid B) = -\log \max_{a,b}\Pr(A=a\mid B=b) =
\min_b \hinf(A\mid B=b)$ (an $\eps$-smooth version is defined
analogously by considering all distributions $(C, D)$ that are within
$\eps$ of $(A, B)$ and taking the maximum among them).  This
definition is too strict: it takes the worst-case $b$, while for
randomness extraction (and many other settings, such as predictability
by an adversary), average-case $b$ suffices.  Average min-entropy
leads to more extractable bits. Nevertheless, after smoothing the two
notions are equivalent up to an additive $\logeps$ term:
$\thepsinf(A\mid B) \ge \hepsinf(A\mid B)$ and
$\hinf^{\eps+\eps_2}(A\mid B) \ge \thepsinf(A\mid B) -\logepstwo$ (for
the case of $\eps=0$, this follows by constructing a new distribution
that eliminates all $b$ for which $\hinf(A\mid B=b) < \thinf(A\mid
B)-\logepstwo$, which will be within $\eps_2$ of the $(A, B)$ by
Markov's inequality; for $\eps>0$, an analogous proof works).  Note
that by~\lemref{hinf}(b), this implies a simple chain rule for
$\hepsinf$ (a more general one is given in~\cite[Section 2.4]{RW05}):
$\hinf^{\eps+\eps_2}(A\mid B) \ge \thepsinf((A, B)) -
H_0(B)-\logepstwo$, where $H_0(B)$ is the logarithm of the number of
possible values of $B$.

\section{Lower Bounds from Coding}
\label{sec:lb}

Recall that an $(\M,K,t)$ code is a subset of the metric space $\M$ which can
\emph{correct} $t$ errors (this is slightly different from the usual notation
of coding theory literature).

Let $K(\M,t)$ be the largest $K$ for which there exists an $(\M,K,t)$-code.
Given any set $S$ of $2^m$ points in $\M$, we let $K(\M,t,S)$ be the largest
$K$ such that there exists an $(\M,K,t)$-code all of whose $K$ points belong to
$S$. Finally, we let $L(\M,t,m) = \log(\min_{|S|=2^m} K(n,t,S))$. Of course,
when $m = \log |\M|$, we get \ifnum\siam=1 that \fi $L(\M,t,n)
=\allowbreak  \log K(\M,t)$. The exact
determination of quantities $K(\M,t)$ and $K(\M,t,S)$ is a central problem of
coding theory and is typically very hard. To the best of our knowledge, the
quantity $L(\M,t,m)$ was not explicitly studied in any of three metrics that we
study, and its exact determination seems hard as well.

We give two simple lower bounds on the entropy loss (one for secure
sketches, the other for fuzzy extractors) which show that our
constructions for the Hamming and set difference metrics output as
much entropy $m'$ as possible when the original input distribution is
uniform. In particular, because the constructions have the same
entropy loss regardless of $m$, they are optimal in terms of the
entropy loss $m-m'$. We conjecture that the constructions also have
the highest possible value $m'$ for all values of $m$, but we do not
have a good enough understanding of $L(\M,t,m)$ (where $\M$ is the
Hamming metric) to substantiate the conjecture.

\begin{lemma}\label{lem:ff-lower}
The existence of an $(\M,m,m',t)$ secure sketch implies that $m' \le
L(\M,t,m)$. In particular, when $m = \log |\M|$ (i.e., when the password
is truly uniform), $m' \le \log K(\M,t)$.
\end{lemma}
\begin{myproof}
Assume $\ff$ is such a secure sketch. Let $S$ be any set of size
$2^m$ in $\M$, and let $W$ be uniform over $S$. Then we must have
$\thinf(W\mid \ff(W)) \ge m'$. In particular, there must be some
value $v$ such that $\hinf(W\mid \ff(W)=v)\ge m'$. But this
means that conditioned on $\ff(W) = v$, there are at least $2^{m'}$
points $w$ in $S$ (call this set $T$) which could produce $\ff(W) =
v$. We claim that these $2^{m'}$ values of $w$ form a code of
error-correcting distance $t$.
Indeed, otherwise there would be a point $w'\in \M$ such that
$\dis{w_0}{w'}\le t$ and $\dis{w_1}{w'}\le t$ for some $w_0,w_1\in T$. But then
we must have that $\Rec(w',v)$ is equal to both $w_0$ and $w_1$, which
is impossible. Thus, the set $T$ above must form an
$(\M,2^{m'},t)$-code inside $S$, which means that $m'\le \log
K(\M,t,S)$. Since $S$ was arbitrary, the bound follows.
\ifnum\siam=1\qquad\fi
\end{myproof}

\begin{lemma}\label{lem:fe-lower}
The existence of $(\M,m,\ell,t,\eps)$-fuzzy extractors implies that
$\ell \le L(\M,t,m) - \log(1-\eps)$. In particular, when $m = \log |\M|$
(i.e., when the password is truly uniform), $\ell \le \log K(\M,t) -
\log(1-\eps)$.
\end{lemma}
\begin{myproof}
Assume $(\Gen, \Rep)$ is such a fuzzy extractor. Let $S$ be any set of
size $2^m$ in $\M$, let $W$ be uniform over $S$ and let $(R,P) \from \Gen(W)$. Then we must have
$\sd{\tuple{R,P}, \tuple{U_\ell,P}} \le \eps$. In particular, there must
be some value $p$ of $P$ such that $R$ is $\eps$-close to
$U_\ell$ conditioned on $P=p$. In particular, this means that
conditioned on $P = p$, there are at least $(1-\eps) 2^{\ell}$ points
$r\in \zo^{\ell}$ (call this set $T$) which could be extracted with
$P=p$. Now, map every $r\in T$ to some arbitrary $w\in S$ which could
have produced $r$ with nonzero probability given $P=p$, and call this
map $C$. $C$ must define a code with error-correcting
distance $t$ by the same reasoning as in \lemref{ff-lower}.
\ifnum\siam=1\qquad\fi
\end{myproof}

Observe that, as long as $\eps<1/2$, we have $0<-\log(1-\eps)<1$, so the
lower bounds on secure sketches and fuzzy extractors differ by less than
a bit.

\section{Analysis of the Original Juels-Sudan Construction}
\label{sec:ojs}
In this section we present a new analysis for the Juels-Sudan secure sketch
for set difference.  We will assume that $n=|\U|$ is a prime power and work
over the field $\F = \GF(n)$. On input set $w$, the original Juels-Sudan
sketch is a list of $r$ pairs of points $(x_i, y_i)$ in $\F$, for some
parameter $r$, $s<r\leq n$.
It is computed as follows:

\begin{constr}[Original Juels-Sudan Secure Sketch \cite{JS02}]
\label{constr:orig-js} \ifnum\siam=0 \ \\ \fi
Input: a set $w\subseteq \F$ of size $s$ and parameters $r\in \set{s+1,\dots,n}, t\in\set{1,\dots,s}$

\ifnum\siam=0
\begin{CompactEnumerate}
\else
\begin{remunerate}
\fi
\item Choose $p()$ at random from the set of polynomials of degree at most
  $k=s-t-1$ over $\F$.
  \\
  Write $w=\set{x_1,\dots,x_s}$, and let $y_i=p(x_i)$ for $i=1,\dots,s$.
\item Choose $r-s$ distinct points $x_{s+1},\dots,x_r$ at random from $\F-w$.
\item For $i=s+1,\dots,r$, choose $y_i\in \F$ at random such that $y_i\neq
  p(x_i)$.
\item Output $\ff(w)=\set{(x_1,y_1),\dots,(x_r,y_r)}$ (in lexicographic order of
  $x_i$).
\ifnum\siam=0
\end{CompactEnumerate}
\else
\end{remunerate}
\fi
\end{constr}

The parameter $t$ measures the error-tolerance of the scheme: given
$\ff(w)$ and a set $w'$ such that $w \triangle w' \leq t$, one can
recover $w$ by considering the pairs $(x_{i},y_{i})$ for $x_{i}\in w'$
and running Reed-Solomon decoding to recover the low-degree polynomial
$p(\cdot)$. When the parameter $r$ is very small, the scheme corrects
approximately twice as many errors with good probability (in the
``input-dependent'' sense from \secref{improved-param-list}). When $r$
is low, however, we show here that the bound on the entropy loss
becomes very weak.

The parameter $r$ dictates the amount of storage necessary, one on
hand, and also the security of the scheme (that is, for $r=s$ the
scheme leaks all information and for larger and larger $r$ there is
less information about $w$).  Juels and Sudan actually propose two
analyses for the scheme. First, they analyze the case where the secret
$w$ is distributed uniformly over all subsets of size $s$. Second,
they provide an analysis of a nonuniform password distribution, but
only for the case $r=n$ (that is, their analysis applies only in the
small universe setting, where $\Omega(n)$ storage is acceptable). Here
we give a simpler analysis which handles nonuniformity and any $r\leq
n$.  We get the same results for a broader set of parameters.

\begin{lemma}
  The entropy loss of the Juels-Sudan scheme is at most $t\log n
  + \log{\binom{n}  r} - \log{\binom{n-s}{r-s}} + 2$.
\end{lemma}

\begin{proof}
This is a simple application of~\lemref{hinf}(b).  $\hinf((W, \ff(W)))$
can be computed as follows.  Choosing the polynomial $p$ (which can be
uniquely recovered from $w$ and $\ff(w)$) requires $s-t$
random choices from $\F$. The choice of the remaining $x_i$'s
requires $\log{\binom{n-s}{r-s}}$ bits, and choosing the $y_i's$
requires $r-s$ random choices from $\F-\set{p(x_i)}$.
Thus,
$\hinf((W,\ff(W))) = \hinf(W)  + (s-t)\log n +
  \log{\binom{n-s}{r-s}}
  + (r-s)\log (n-1)$.
The output can be described in $\log\paren{{\binom{n}  r} n^r}$
bits. The result follows by \lemref{hinf}(b) after observing that
$(r-s)\log\frac{n}{n-1} < n \log \frac{n}{n-1} \le 2$.
\ifnum\siam=1\qquad\fi
\end{proof}
In the large universe setting, we will have $r \ll n$ (since we wish
to have storage polynomial in $s$). In that setting, the bound on
the entropy loss of the Juels-Sudan scheme is in fact very large. We
can
rewrite the entropy loss as $t\log n - \log {\binom{r}{s}} +
\log {\binom{n}{s}} + 2$, using the identity ${\binom{n}{r}} {\binom r
s} = {\binom n s}{\binom{n-s} { r-s}}$. Now the entropy of $W$ is at
most ${\binom n s}$, and so our lower bound on the remaining entropy
is $(\log{\binom r s} - t\log n - 2)$. To make this quantity large
requires making $r$ very large.

\section{BCH Syndrome Decoding in Sublinear Time}
\label{sec:bch}

We show that the standard decoding algorithm for BCH codes can be
modified to run in time polynomial in the length of the syndrome. This works
for BCH codes over any field $\GF(q)$, which include Hamming codes in
the binary case and Reed-Solomon for the case $n = q-1$. BCH codes are
handled in detail in many textbooks (e.g., \cite{vanLint}); our
presentation here is quite terse. For simplicity, we discuss only
primitive, narrow-sense BCH codes here; the discussion extends easily
to the general case.

The algorithm discussed here has been revised due to an error pointed
out by Ari Trachtenberg.  Its implementation is available~\cite{ijs-bch-impl}.

We'll use a slightly nonstandard formulation of BCH codes.  Let
$n=q^m-1$ (in the binary case of interest in \secref{sublin}, $q=2$).
We will work in two finite fields: $\GF(q)$ and a larger extension
field $\F=\GF(q^m)$.  BCH codewords, formally defined below, are then
vectors in $\GF(q)^n$.  In most common presentations, one indexes the
$n$ positions of these vectors by discrete logarithms of the elements
of $\F^*$: position $i$, for $1\le i \le n$, corresponds to
$\alpha^i$, where $\alpha$ generates the multiplicative group $\F^*$.
However, there is no inherent reason to do so: they can be indexed by
elements of $\F$ directly rather than by their discrete logarithms.
Thus, we say that a word has value $p_x$ at position $x$, where $x\in
\F^*$.  If one ever needs to write down the entire $n$-character word
in an ordered fashion, one can arbitrarily choose a convenient
ordering of the elements of $\F$ (e.g., by using some standard binary
representation of field elements); for our purposes this is not
necessary, as we do not store entire $n$-bit words explicitly, but
rather represent them by their supports: $\supp(v)=\{(x, p_x)\mid
p_x\neq 0\}$.  Note that for the binary case of interest in
\secref{sublin}, we can define $\supp(v)=\{x\mid p_x \neq 0\}$,
because $p_x$ can take only two values: 0 or 1.

Our choice of representation will be crucial for efficient decoding:
in the more common representation, the last step of the decoding
algorithm requires one to find the position $i$ of the error from the
field element $\alpha^i$.  However, no efficient algorithms for
computing the discrete logarithm are known if $q^m$ is large (indeed, a
lot of cryptography is based on the assumption that such an efficient
algorithm does not exist).  In our representation, the field element
$\alpha^i$ will in fact be the position of the error.

\begin{defn}\label{def:bch2}
    The (narrow-sense, primitive) BCH code of designed distance $\delta$ over $\GF(q)$ (of length $n\ge \delta$) is given by the set of vectors of the form $\big(c_x\big)_{x \in \F^*}$ such that each $c_x$ is in the smaller field $\GF(q)$, and the vector satisfies the constraints $\sum_{x \in \F^*} c_x x^i = 0$, for $i=1,\ldots,\delta -1$, with arithmetic done in the larger field $\F$.
\end{defn}

\newcommand{\dlog}{\mathsf{dlog}}
\newcommand{\lex}{\mathsf{lex}}
\newcommand{\numerrors}{e}

To explain this definition, let us fix a generator $\alpha$ of the multiplicative group of the large field $\F^*$. For any vector of coefficients $\big(c_x\big)_{x \in \F^*}$, we can define a polynomial
$$c(z) = \sum_{x\in \GF(q^m)^*} c_x z^{\dlog(x)}\,,$$
where $\dlog(x)$ is the discrete logarithm of $x$ with respect to
$\alpha$.  The conditions of the definition are then equivalent to the
requirement (more commonly seen in presentations of BCH codes) that
$c(\alpha^i)= 0$ for $i=1,\ldots,\delta-1$, because
$(\alpha^i)^{\dlog(x)} = (\alpha^{\dlog(x)})^i = x^i$.

We can simplify this somewhat. Because the coefficients $c_x$ are in
$\GF(q)$, they satisfy $c_x^q=c_x$. Using the identity $(x+y)^q=x^q + y^q$,
which holds even in the large field $\F$, we have $c(\alpha^{i})^q =
\sum_{x\neq 0} c_x^q x^{iq} = c(\alpha^{iq})$. Thus, roughly a $1/q$
fraction of the conditions in the definition are redundant: we need only to
check that they hold for $i\in\set{1,\dots,\delta-1}$ such that $q\not | i$.

The syndrome of a word (not necessarily a codeword) $ (p_x)_{x\in \F^*} \in \GF(q)^n$ with respect to the BCH code above is the vector
$$\syn(p) = p(\alpha^1),\ldots, p(\alpha^{\delta-1}), \quad  \text{where} \quad p(\alpha^i) = \sum_{x\in\F^*} p_x x^i.$$
As mentioned above, we do not in fact have to include the values $p(\alpha^i)$ such that $q|i$.

\mypar{Computing with Low-Weight Words}
A low-weight word $p\in \GF(q)^n$ can be represented either as a long
string or, more compactly, as a list of positions where it is nonzero
and its values at those points. We call this representation the
support list of $p$ and denote it $\supp(p) = \set{(x,p_x)}_{x:
  p_x\neq 0}$.

\begin{lemma}\label{lem:bchgen}
  For a $q$-ary BCH code $C$ of designed distance $\delta$, one can
  compute:
  \ifnum\siam=0
  \begin{CompactEnumerate}
  \else
  \begin{remunerate}
  \fi
    \item $\syn(p)$ from $\supp(p)$ in time polynomial in $\delta$,
    $\log n$, and $|\supp(p)|$, and
    \item $\supp(p)$ from $\syn(p)$ (when $p$ has weight at most
      $(\delta -1)/2$), in time polynomial in $\delta$ and $\log n$.
  \ifnum\siam=0
  \end{CompactEnumerate}
  \else
  \end{remunerate}
  \fi
\end{lemma}

\begin{proof}
  Recall that $\syn(p) = (p(\alpha),\dots,p(\alpha^{\delta-1}))$ where
  $p(\alpha^i) = \sum_{x\neq 0} p_x x^i$.  Part (1) is easy, since to
compute the syndrome we need only to compute the powers of $x$. This requires about $\delta\cdot \weight(p)$ multiplications in $\F$. For
  Part (2), we adapt Berlekamp's BCH
  decoding algorithm, based on its presentation in \cite{vanLint}. Let
  $M=\set{x \in \F^*| p_x \neq 0}$, and define
  $$\sigma(z) \defeq \prod_{x\in M} (1- x z) \quad \mbox{and}\quad
  \omega(z) \defeq \sigma(z) \sum_{x\in M} \frac{p_x x z}
  {(1-x z)}\,.$$
  Since $(1-x z)$ divides $\sigma(z)$ for
  $x\in M$, we see that $\omega(z)$ is in fact a polynomial of degree
  at most $|M|=\weight(p)\leq (\delta-1)/2$.  The polynomials
  $\sigma(z)$ and $\omega(z)$ are known as the error locator
  polynomial and evaluator polynomial, respectively; observe that
  $\gcd(\sigma(z), \omega(z))=1$.

  We will in fact work with our polynomials modulo $z^\delta$. In this  arithmetic the inverse of $(1-xz)$ is $\sum_{\ell=1}^{\delta} (xz)^{\ell-1}$; that is,
$$(1-xz)\sum_{\ell=1}^{\delta} (xz)^{\ell-1} \equiv  1 \mod z^\delta.$$
We are given $p(\alpha^\ell)$ for $\ell = 1,\dots,\delta$.
  Let $S(z) =\sum_{\ell=1}^{\delta-1}p(\alpha^\ell) z^\ell$. Note that $S(z) \equiv \sum_{x\in M} p_x \frac{xz}{(1-xz)}\mod z^\delta$. This implies that
  $$S(z)\sigma(z) \equiv \omega(z) \mod {z^\delta}.$$

  The polynomials $\sigma(z)$ and $\omega(z)$ satisfy the following four
  conditions: they are of degree at most $(\delta-1)/2$ each, they are
  relatively prime, the constant coefficient of $\sigma$ is 1, and they
  satisfy this congruence.  In fact, let $w'(z),\sigma'(z)$ be any nonzero
  solution to this congruence, where degrees of $w'(z)$ and $\sigma'(z)$ are
  at most $(\delta-1)/2$. Then $w'(z)/\sigma'(z) =
  \omega(z)/\sigma(z)$. (To see why this is so, multiply the initial
  congruence by $\sigma'()$ to get $\omega(z)\sigma'(z) \equiv
  \sigma(z)\omega'(z) \mod z^\delta$.  Since both sides of the
  congruence have degree at most $\delta -1$, they are in fact equal as
  polynomials.)  Thus, there is at most one solution $\sigma(z), \omega(z)$
  satisfying all four conditions, which can be obtained from any
  $\sigma'(z),\omega'(z)$ by reducing the resulting fraction
  $\omega'(z)/\sigma'(z)$ to obtain the  solution of
  minimal degree with the constant term of $\sigma$ equal to 1.

  Finally, the roots of $\sigma(z)$ are the points $x^{-1}$ for $x\in M$,
  and the exact value of $p_x$ can be recovered from
  $\omega(x^{-1}) = p_x \prod_{y\in M, y\neq x} (1-yx^{-1})$ (this is
  needed only for $q>2$, because for $q=2$, $p_x=1$).  Note that it is possible
  that a solution to the congruence will be found even if the input
  syndrome is not a syndrome of any $p$ with $\weight(p)>(\delta-1)/2$ (it
  is also possible that a solution to the congruence will not be found at
  all, or that the resulting $\sigma(z)$ will not split into distinct nonzero
  roots).  Such a solution will not give the correct $p$.  Thus, if there
  is no guarantee that $\weight(p)$ is actually at most $(\delta-1)/2$, it
  is necessary to recompute $\syn(p)$ after finding the solution, in order
  to verify that $p$ is indeed correct.

  Representing coefficients of $\sigma'(z)$ and $\omega'(z)$ as unknowns, we
  see that solving the congruence requires only solving a system of $\delta$
  linear equations (one for each degree of $z$, from 0 to $\delta-1$)
  involving $\delta+1$ variables over $\F$, which can
  be done in $O(\delta^3)$ operations in $\F$ using, e.g., Gaussian
  elimination.  The reduction of the fraction $\omega'(z)/\sigma'(z)$
  requires simply running Euclid's algorithm for finding the g.c.d. of
  two polynomials of degree less than $\delta$, which takes
  $O(\delta^2)$ operations in $\F$.  Suppose the resulting $\sigma$
  has degree $\numerrors$.  Then one can find the roots of $\sigma$ as
  follows.  First test that $\sigma$ indeed has $\numerrors$ distinct
  roots by testing that $\sigma(z) | z^{q^m}-z$ (this is a necessary
  and sufficient condition, because every element of $\F$ is a root of
  $z^{q^m}-z$ exactly once).  This can be done by computing $(z^{q^m}
  \bmod \sigma(z))$ and testing if it equals $z\bmod \sigma$; it takes
  $m$ exponentiations of a polynomial to the power $q$, i.e., $O((m
  \log q) \numerrors^2)$ operations in $\F$.  Then apply an
  equal-degree-factorization algorithm (e.g., as described
  in~\cite{Sho05}), which also takes $O((m \log q) \numerrors^2)$
  operations in $\F$.  Finally, after taking inverses of the roots of
  $\F$ and finding $p_x$ (which takes $O(\numerrors^2)$ operations in $\F$),
  recompute $\syn(p)$ to verify that it is equal to the input value.

  Because $m \log q = \log (n+1)$ and $\numerrors \le (\delta-1)/2$, the total
  running time is $O(\delta^3 + \delta^2 \log n)$ operations in $\F$;
  each operation in $\F$ can done in time $O(\log^2 n)$, or faster
  using advanced techniques.

One can improve this running time substantially. The error locator
polynomial $\sigma()$ can be found in $O(\log \delta)$ convolutions
(multiplications) of polynomials over $\F$ of degree $(\delta-1)/2$
each~\cite[Section 11.7]{Blahut83} by exploiting the special structure
of the system of linear equations being solved.  Each convolution can
be performed asymptotically in time $O(\delta \log \delta \log \log
\delta)$ (see, e.g.,~\cite{vzGG03}), and the total time required to find
$\sigma$ gets reduced to $O(\delta \log^2 \delta \log \log \delta)$
operation in $\F$.  This replaces the $\delta^3$ term in the above
running time.

While this is asymptotically very good, Euclidean-algorithm-based
decoding~\cite{SKHN75}, which runs in $O(\delta^2)$
operations in $\F$, will find $\sigma(z)$ faster for reasonable values
of $\delta$ (certainly for $\delta<1000)$.  The algorithm finds
$\sigma$ as follows:
{\tt
\newcommand{\Rold}{{R_{\mathrm{old}}}}
\newcommand{\Rnew}{{R_{\mathrm{new}}}}
\newcommand{\Rcur}{{R_{\mathrm{cur}}}}
\newcommand{\Vold}{{V_{\mathrm{old}}}}
\newcommand{\Vnew}{{V_{\mathrm{new}}}}
\newcommand{\Vcur}{{V_{\mathrm{cur}}}}
\begin{tabbing}
123\=12345\=123\=\kill
\> set $\Rold(z) \gets z^{\delta-1}$, $\Rcur(z) \gets S(z)/z$,
$\Vold(z)\gets 0$, $\Vcur(z) \gets 1$.\\
\>while $\deg(\Rcur(z)) \ge (\delta-1)/2$: \\
\> \>divide $\Rold(z)$ by $\Rcur(z)$ to get quotient $q(z)$ and remainder
\ifnum\siam=1 \\ \>\>\> \fi
$\Rnew(z)$;\\
\>\> set $\Vnew(z) \gets \Vold(z) - q(z)\Vcur(z)$;\\
\>\> set $\Rold(z) \gets \Rcur(z), \Rcur(z) \gets \Rnew(z), \Vold(z)
\gets \Vcur(z),$ \ifnum\siam=1\\ \>\>\>\fi $\Vcur(z) \gets \Vnew(z)$.\\
\> set $c \gets \Vcur(0)$; set $\sigma(z) \gets \Vcur(z)/c$ and $\omega(z) \gets z\cdot \Rcur(z)/c$
\end{tabbing}
}
\ifnum\siam=0\noindent\fi
In the above algorithm, if $c=0$, then
the correct $\sigma(z)$ does not exist,
i.e., $\weight(p) > (\delta-1)/2$.
The correctness of this algorithm can be seen by observing
that the congruence $S(z)\sigma(z) \equiv \omega(z) \pmod {z^\delta}$ can
have $z$ factored out of it (because $S(z)$, $\omega(z)$ and $z^\delta$
are all divisible by $z$) and rewritten as $(S(z)/z)\sigma(z) + u(z)
z^{\delta-1} = \omega(z)/z$, for some $u(z)$.  The obtained $\sigma$ is
easily
shown to be the correct one (if one exists at all) by
applying~\cite[Theorem
18.7]{Sho05} (to use the notation of that theorem, set $n=z^{\delta-1},
y=S(z)/z, t^*=r^* = (\delta-1)/2, r'=\omega(z)/z, s'=u(z), t'=\sigma(z)$).

The root finding of $\sigma$ can also be sped up.  Asymptotically,
detecting if a polynomial over $\F=\GF(q^m)=\GF(n+1)$ of degree $\numerrors$
has $\numerrors$ distinct roots and finding these roots can be
performed in time $O(\numerrors^{1.815} (\log n)^{0.407})$ operations in $\F$
using the algorithm
of Kaltofen and Shoup~\cite{KF95}, or in time
$O(\numerrors^2 + (\log n) \numerrors \log \numerrors
\log \log \numerrors)$ operations in $\F$ using the EDF algorithm
of Cantor and Zassenhaus\footnote{See~\cite[Section
21.3]{Sho05}, and substitute the most efficient known polynomial
arithmetic. For example, the procedures described in~\cite{vzGG03} take time
$O(\numerrors \log \numerrors \log \log \numerrors)$ instead of time
$O(\numerrors^2)$  to perform modular arithmetic operations with
degree-$\numerrors$ polynomials.}. For reasonable values of
$\numerrors$, the Cantor-Zassenhaus EDF algorithm with
Karatsuba's multiplication algorithm~\cite{KO63} for polynomials
will be
faster, giving root-finding running time of
$O(\numerrors^2+\numerrors^{\log_2 3} \log n)$ operations in $\F$.
Note that if the actual weight $\numerrors$ of $p$ is close to the maximum
tolerated $(\delta-1)/2$, then finding the roots of $\sigma$ will
actually take longer than finding $\sigma$.
\ifnum\siam=1\qquad\fi
\end{proof}

\mypar{A Dual View of the Algorithm}
Readers may be \ifnum\siam=0 used to seeing \else more familiar with \fi a different, evaluation-based formulation of BCH
codes, in which codewords are generated as follows.  Let $\F$ again be an
extension of $\GF(q)$, and let $n$ be the length of the code (note that
$|\F^*|$ is not necessarily equal to $n$ in this formulation).  Fix distinct $x_1, x_2,
\dots, x_n \in \F$.  For every polynomial $c$ over the large field $\F$ of
degree at most $n-\delta$, the vector $(c(x_1), c(x_2), \dots c(x_n))$ is a
codeword if and only if every coordinate of the vector happens to be in the
smaller field: $c(x_i)\in \GF(q)$ for all $i$.  In particular, when
$\F=\GF(q)$, then every polynomial leads to a codeword, thus giving
Reed-Solomon codes.

The syndrome in this formulation can be computed as follows: given a vector
$y=(y_1, y_2, \dots, y_n)$ find the interpolating polynomial $P=p_{n-1}
x^{n-1} + p_{n-2} x^{n-2} + \dots + p_0$ over $\F$ of degree at most $n-1$
such that $P(x_i)=y_i$ for all $i$.  The syndrome is then the negative top
$\delta-1$ coefficients of $P$: $\syn(y)=(-p_{n-1}, -p_{n-2}, \dots,
-p_{n-(\delta-1)})$.  (It is easy to see that this is a syndrome: it is a
linear function that is zero exactly on the codewords.)

When $n=|\F|-1$, we can index the $n$-component vectors by elements of
$\F^*$, writing codewords as $(c(x))_{x\in \F^*}$.  In this case, the
syndrome of $(y_x)_{x\in \F^*}$ defined as the negative top $\delta-1$
coefficients of $P$ such that for all $x\in F^*, P(x)=y_x$ is equal to the
syndrome defined following \defref{bch2} as $\sum_{x\in \F} y_x x^i$ for
$i=1, 2, \dots, \delta-1$. \footnote{
This statement can be shown as follows:
because both maps are linear, it is sufficient to prove that they agree on
a vector $(y_x)_{x\in \F^*}$ such that $y_a=1$ for some $a\in
\F^*$ and $y_x=0$ for $x\ne a$.  For such a vector, $\sum_{x\in \F} y_x
x^i = a^i$.  On the other hand, the interpolating polynomial $P(x)$ such
that $P(x)=y_x$ is $-a x^{n-1} - a^2 x^{n-2} - \dots - a^{n-1} x - 1$
(indeed, $P(a)=-n=1$; furthermore, multiplying $P(x)$ by $x-a$ gives $a(x^n
-1)$, which is zero on all of $\F^*$; hence $P(x)$ is zero for every $x\ne
a$).}
Thus, when $n=|\F|-1$, the codewords obtained via the evaluation-based
definition are \emph{identical} to the codewords obtain via \defref{bch2},
because codewords are simply elements with the zero syndrome, and the
syndrome maps agree.

This is an example of a remarkable duality between evaluations of
polynomials and their coefficients: the syndrome can be viewed either as the
evaluation of a polynomial whose coefficients are given by the vector, or
as the coefficients of the polynomial whose evaluations are given by a
vector.

The syndrome decoding algorithm above has a natural interpretation in the
evaluation-based view.  Our presentation is an adaptation of
Welch-Berlekamp decoding as presented in, e.g.,~\cite[Chapter 10]{Sud01}.

Suppose $n=|F|-1$ and  $x_1,\dots,x_n$ are the nonzero elements of the field.
Let $y=(y_1, y_2, \dots, y_n)$ be a vector.  We are given its syndrome
$\syn(y)=(-p_{n-1}, -p_{n-2}, \dots,\allowbreak -p_{n-(\delta-1)})$, where $p_{n-1},
\dots, p_{n-(\delta-1)}$ are the top coefficients of the interpolating
polynomial $P$.  Knowing only $\syn(y)$, we need to find at most
$(\delta-1)/2$ locations $x_i$ such that correcting all the corresponding
$y_i$ will result in a codeword.  Suppose that codeword is given by a
degree-$(n-\delta)$ polynomial $c$. Note that $c$ agrees with $P$ on all
but the error locations.  Let $\rho(z)$ be the polynomial of degree at most
$(\delta-1)/2$ whose roots are exactly the error locations.  (Note that
$\sigma(z)$ from the decoding algorithm above is the same $\rho(z)$ but
with coefficients in reverse order, because the roots of $\sigma$ are the
inverses of the roots of $\rho$.) Then $\rho(z) \cdot P(z) = \rho(z) \cdot
c(z)$ for $z=x_1, x_2,
\dots, x_n$. Since $x_1,\dots,x_n$ are all the nonzero field elements, $\prod_{i=1}^n (z-x_i) = z^n-1$. Thus,
\begin{equation*}
\rho(z) \cdot c(z)\quad  =\quad  \rho(z) \cdot P(z) \bmod
\prod_{i=1}^n (z-x_i)\quad  =\quad \rho(z) \cdot P(z) \bmod
(z^n-1)
\,.
\end{equation*}

If we write the left-hand side as $\alpha_{n-1} x^{n-1}+\alpha_{n-2}
x^{n-2} + \cdots + \alpha_0$, then the above equation implies that
$\alpha_{n-1}= \cdots =\alpha_{n-(\delta-1)/2}=0$ (because the
degree if $\rho(z)\cdot c(z)$ is at most $n-(\delta+1)/2$).  Because
$\alpha_{n-1}, \dots, \alpha_{n-(\delta-1)/2}$ depend on the
coefficients of $\rho$ as well as on $p_{n-1}, \dots,
p_{n-(\delta-1)}$, but not on lower coefficients of $P$, we obtain a
system of $(\delta-1)/2$ equations for $(\delta-1)/2$ unknown
coefficients of $\rho$. A careful examination shows that it is
essentially the same system as we had for $\sigma(z)$ in the
algorithm above.  The lowest-degree solution to this system is
indeed the correct $\rho$, by the same argument which was used to
prove the correctness of $\sigma$ in \lemref{bchgen}. The roots of
$\rho$ are the error-locations. For $q>2$, the actual corrections
that are needed at the error locations (in other words, the light
vector corresponding to the given syndrome) can then be recovered by
solving the linear system of equations implied by the value of the
syndrome.

\ifnum\siam=1
\textbf{Acknowledgments.\ }
This work evolved over several years, and discussions with many people
enriched our understanding of the material at hand. In roughly
chronological order, we thank Piotr Indyk for discussions about embeddings
and for his help in the proof of~\lemref{edit}; Madhu Sudan for helpful
discussions about the construction of \cite{JS02} and the uses of
error-correcting codes; Venkat Guruswami for enlightenment about list
decoding; Pim Tuyls for pointing out relevant previous work; Chris
Peikert for pointing out the model of computationally bounded adversaries
from \cite{MPSW05}; Ari Trachtenberg for finding an error in the
preliminary version of \appref{bch}; Ronny Roth for discussions about
efficient BCH decoding; Kevin Harmon and Soren Johnson for their
implementation work; and Silvio Micali and anonymous referees
for suggestions on presenting our
results.

The work of the Y.D. was partly funded by the National Science
Foundation under CAREER Award No. CCR-0133806 and Trusted Computing
Grant No. CCR-0311095, and by the New York University Research
Challenge Fund 25-74100-N5237.  The work of the L.R. was partly funded
by the National Science Foundation under Grant Nos. CCR-0311485, CCF-0515100 and CNS-0202067.  The
work of the A.S. at MIT was partly funded by US A.R.O. grant
DAAD19-00-1-0177 and by a Microsoft Fellowship. While at the Weizmann Institute,
A.S. was supported by the Louis L. and Anita M. Perlman Postdoctoral
Fellowship.

\bibliographystyle{siam}
\bibliography{fuzzy}
\fi
\end{document}